\documentclass[fleqn,usenatbib]{mnras}

\usepackage{newtxtext,newtxmath}

\usepackage[T1]{fontenc}

\DeclareRobustCommand{\VAN}[3]{#2}
\let\VANthebibliography\thebibliography
\def\thebibliography{\DeclareRobustCommand{\VAN}[3]{##3}\VANthebibliography}

\usepackage{graphicx}	
\usepackage{amsmath}

\title{The 2022--2023 accretion outburst of the young star V1741 Sgr}

\author[M. A. Kuhn et al.]{Michael A. Kuhn,$^{1}$\thanks{E-mail: m.kuhn@herts.ac.uk}
Lynne A. Hillenbrand,$^{2}$
Michael S. Connelley,$^{3}$
R.\ Michael Rich,$^4$
Bart Staels,$^5$ \newauthor
Adolfo S. Carvalho,$^{2}$
Philip W. Lucas,$^{1}$
Christoffer Fremling$,^{2}$
Viraj R. Karambelkar,$^{2}$
Ellen Lee,$^{3}$ \newauthor
Tom\'as Ahumada$,^{2}$
Emille E. O. Ishida,$^{6}$
Kishalay De,$^{7}$
Rafael S. de Souza,$^{1}$
Mansi Kasliwal$^{2}$
\\
$^{1}$Centre for Astrophysics Research, Department of Physics, Astronomy and Mathematics, University of Hertfordshire, Hatfield, AL10 9AB, UK\\
$^{2}$Department of Astronomy, California Institute of Technology, 1216 East California Boulevard, Pasadena, CA 91125, USA\\
$^{3}$University of Hawaii at Manoa, Institute for Astronomy, 640 North Aohoku Place, Hilo, HI 96720, USA\\
$^{4}$Department of Physics \& Astronomy, University of California Los Angeles, PAB 430 Portola Plaza, Los Angeles, CA 90095-1547, USA\\
$^{5}$American Association of Variable Star Observers\\
$^{6}$Universit\'e Clermont Auvergne, CNRS/IN2P3, LPC, F-63000 Clermont-Ferrand, France\\
$^{7}$Kavli Institute for Astrophysics and Space Research, Massachusetts Institute of Technology, Cambridge, MA, USA\\
}

\date{Accepted 2024 January 11. Received 2024 January 11; in original form 2023 July 10}

\pubyear{2023}

\begin{document}
\label{firstpage}
\pagerange{\pageref{firstpage}--\pageref{lastpage}}
\maketitle

\begin{abstract}
V1741 Sgr (= SPICY 71482/Gaia22dtk) is a Classical T Tauri star on the outskirts of the Lagoon Nebula. After at least a decade of stability, in mid-2022, the optical source brightened by $\sim$3~mag over two months, remained bright until early 2023, then dimmed erratically over the next four months. This event was monitored with optical and infrared spectroscopy and photometry. Spectra from the peak (October 2022) indicate an EX Lup-type (EXor) accretion outburst, with strong emission from H\,{\sc i}, He\,{\sc i}, and Ca\,{\sc ii} lines and CO bands. At this stage, spectroscopic absorption features indicated a temperature of $T\sim4750$~K with low-gravity lines (e.g., Ba\,{\sc ii} and Sr\,{\sc ii}). By April 2023, with the outburst beginning to dim, strong TiO absorption appeared, indicating a cooler $T\sim3600$~K temperature. However, once the source had returned to its pre-outburst flux in August 2023, the TiO absorption and the CO emission disappeared. When the star went into outburst, the source's spectral energy distribution became flatter, leading to bluer colours at wavelengths shorter than $\sim$1.6~$\upmu$m and redder colours at longer wavelengths. The brightening requires a continuum emitting area larger than the stellar surface, likely from optically thick circumstellar gas with cooler surface layers producing the absorption features. Additional contributions to the outburst spectrum may include blue excess from hotspots on the stellar surface, emission lines from diffuse gas, and reprocessed emission from the dust disc. Cooling of the circumstellar gas would explain the appearance of TiO, which subsequently disappeared once this gas had faded and the stellar spectrum reemerged.
\end{abstract}

\begin{keywords}
accretion, accretion discs -- stars: pre-main-sequence -- stars: individual: V1741 Sgr -- stars: variables: T Tauri, Herbig Ae/Be -- techniques: spectroscopic
\end{keywords}

\section{Introduction}


Accretion onto young stellar objects (YSO) leads to a wide range of variable behaviours, from low-amplitude variations on timescales of days to weeks, to outbursts of several magnitudes lasting months to decades. Traditionally, outbursts have been divided into two phenomenological classes: the FU Ori class, whose outbursts last many decades and have a distinct absorption spectral signature, and the EX Lup class (also known as EXors), whose outbursts last months or years and exhibit atomic and molecular emission features \citep{1996ARA&A..34..207H,Audard2014}. Nevertheless, there is considerable diversity in the light curves and spectra of these accretion outbursts, suggesting the possibility of additional classes and/or a behavioural continuum \citep{Fischer2023,Semkov2023}. EX Lup-type stars, in particular, display considerable variation in the strengths of their emission features \citep[e.g.,][]{2022ApJ...929..129G}. Furthermore, some YSO outbursts display mixed FU Ori and EX Lup-like features \citep[e.g.,][]{2017MNRAS.465.3039C,Connelley2018,Hillenbrand2022}, whereas others have nearly featureless spectra \citep[e.g.,][]{2020MNRAS.499.1805L,TWang2023}. 

\begin{table*}
	\centering
	\caption{Spectroscopic Follow-up Log}
	\label{tab:obslog}
	\begin{tabular}{lrllrcrrrlll} 
		\hline
		\multicolumn{1}{c}{Date} & 
		\multicolumn{1}{c}{MJD$^\dagger$} & 
		\multicolumn{1}{c}{Instrument} &
		\multicolumn{1}{c}{Facility} & 
		\multicolumn{1}{c}{Exp} &
		\multicolumn{1}{c}{Airmass} &
		\multicolumn{1}{c}{Slit} &
                \multicolumn{1}{c}{Wavelengths} &
		\multicolumn{1}{c}{$R$} &
		\multicolumn{1}{c}{Standard} &
		\multicolumn{1}{c}{PI} &
		\multicolumn{1}{c}{Stage} \\
		\multicolumn{1}{c}{} & 
		\multicolumn{1}{c}{day} & 
		\multicolumn{1}{c}{} &
		\multicolumn{1}{c}{} & 
		\multicolumn{1}{c}{s} &
		\multicolumn{1}{c}{} &
		\multicolumn{1}{c}{$^{\prime\prime}$} &
                \multicolumn{1}{c}{\AA} &
		\multicolumn{1}{c}{} &
		\multicolumn{1}{c}{} &
		\multicolumn{1}{c}{} &
		\multicolumn{1}{c}{} \\
  \multicolumn{1}{c}{(1)} & 
  \multicolumn{1}{c}{(2)} & 
  \multicolumn{1}{c}{(3)} & 
  \multicolumn{1}{c}{(4)} & 
  \multicolumn{1}{c}{(5)} & 
  \multicolumn{1}{c}{(6)} & 
  \multicolumn{1}{c}{(7)} & 
  \multicolumn{1}{c}{(8)} & 
  \multicolumn{1}{c}{(9)} & 
  \multicolumn{1}{c}{(10)} & 
  \multicolumn{1}{c}{(11)} & 
  \multicolumn{1}{c}{(12)} \\
		\hline
2022 Oct 15 & 59868.2 & SpeX & IRTF &	720 & 1.7 & 0.5 & 7,000--25,000	& $\sim$2000	 		& HD 159415 & M.\ Connelley & Peak\\
2022 Oct 18 & 59871.2 & LRIS-blue & Keck I & 300 &1.6	& 1.0	& 3,100--5,650	& $\sim$900				& G191-B2B & C.\ Fremling & Peak\\
2022 Oct 18 & 59871.2 & LRIS-red & Keck I & 300 &1.6 & 1.0	& 5,650--10,200	& $\sim$900				& BD+28 4211 & C.\ Fremling & Peak\\
2023 Apr 24 &60058.5 & Kast-blue &  Shane & 650 & 2.1 &1.5 & 3,600--5,500 &$\sim$1100			& Feige 34& R. M.\ Rich & Decay\\
2023 Apr 24 & 60058.5 & Kast-red & Shane & 310 & 2.1 &1.5 & 5,300--10,300 &$\sim$1750			& Feige 34& R. M.\ Rich & Decay\\
2023 Aug 5 & 60162.2& TripleSpec & P200 &150 & 1.9 & 1.0 & 10,500--24,500 &$\sim$2600		& HD 157734 & C.\ Fremling & Quiesc.\\
2023 Aug 18 & 60175.2 & Kast-blue & Shane & 1600	& 2.1	&  1.5 & 3,600--5,500 &$\sim$1100		& P330E& R. M.\ Rich & Quiesc.\\
  \hline
\end{tabular}
\flushleft{$^\dagger$Modified Julian day (MJD) is defined as heliocentric Julian day (HJD) minus 2400000.5.}
\flushleft{(The reduced spectra are provided online as supplementary material.)}
\end{table*}

The collection of well-characterised outbursts is still relatively small and only sparsely samples the range of possible outburst behaviours and spectral properties. This poses challenges to refining classifications and, thereby, our physical understanding of these events. Recent discovery rates range from one to several per year, mainly from time-domain surveys monitoring large areas of the sky \citep[e.g.,][]{2021AJ....161..220H,2022ApJ...927..125C,2022ApJ...941..165P,2023JKAS...56..253C,2023MNRAS.521.5669C}. However, historical discovery rates have been much lower. Furthermore, the characterisation of EX Lup-type outbursts at their peak requires relatively quick follow-up spectroscopy, meaning that rapid identification of outbursts is essential. 

Here, we discuss a year-long EX Lup-type outburst from the Classical T Tauri star V1741~Sgr ($18^\mathrm{h}\,02^\mathrm{m}\,14.\mkern-5mu^\mathrm{s}3$ $-24^\circ\, 03^\prime\,47^{\prime\prime}$). This source was first noted as a variable star in the vicinity of the Lagoon Nebula by \citet{Walker1957}.\footnote{V1741~Sgr is entry \#1 in the table of variable stars from \citet[][]{Walker1957} based on observations from the 20-inch Palomar reflector. However, it lies outside the field of view of Herbig's grating plates, so it was not assigned a LkH$\alpha$ number.} However, observers paid it little attention in the subsequent decades. More recently, its infrared excess was detected in Spitzer/Infrared Array Camera \citep[IRAC;][]{2004ApJS..154....1W,2004ApJS..154...10F} photometry from the Galactic Legacy Infrared Midplane Survey Extraordinaire \citep[GLIMPSE;][]{Benjamin2003,Churchwell2009}. Based on these data, the source was included in the Spitzer/IRAC Candidate YSO (SPICY) catalogue (labelled SPICY~71482), where its spectral index $\alpha=-0.7$ suggests a Class~II (disc-bearing) evolutionary stage \citep{SPICY}.

The 2022--2023 outburst was registered in the Gaia Alerts \citep{Hodgkin2021} stream as Gaia22dtk \citep{2022TNSTR2604....1H,2022ATel15721....1K} -- the first recorded outburst from this source. The distinctiveness of this event in the source's light curve makes it a good case study for understanding accretion outbursts from later (post-embedded) evolutionary stages of YSOs.

In this study, we use photometry and spectroscopy to constrain the origin of the increased luminosity of the source as the outburst evolved. Section~\ref{sec:data} describes the data. Section~\ref{sec:outburst} examines the outburst light curve. Section~\ref{sec:spec} investigates the outburst spectra. Section~\ref{sec:evolution} tracks the source's spectral evolution. Section~\ref{sec:star} infers stellar properties. Section~\ref{sec:phys} characterises the source's colour evolution. Section~\ref{sec:discussion} discusses the astrophysical nature of the outbursting YSO. Finally, Section~\ref{sec:conclusion} provides our conclusions.

\section{Data and Observations}\label{sec:data}

The source's light curves and alert packets were obtained from time-domain photometric surveys, including the Zwicky Transient Facility \citep[ZTF;][]{2019PASP..131a8002B,2019PASP..131g8001G}, the Palomar Gattini-Infrared survey \citep[PGIR;][]{2020PASP..132b5001D}, Gaia \citep{GaiaCollaboration2016,Hodgkin2021}, the Visible and Infrared Survey Telescope for Astronomy (VISTA) Variables in the V\'ia L\'actea \citep[VVV;][]{2010NewA...15..433M}, and NEOWISE \citep{Mainzer2011}. Other photometry was obtained from VPHAS+ \citep{2014MNRAS.440.2036D}, Pan-STARRS \citep{Chambers2016,Flewelling2020}, SDSS DR12 \citep{Alam2015}, DENIS \citep{Fouque00}, 2MASS \citep{Skrutskie2006}, UKIDSS \citep{Lawrence2007,Lucas2008}, GLIMPSE, and MIPSGAL \citep{Carey2009,Gutermuth2015} surveys. Finally, the source was monitored during its 2023 dimming by the American Association of Variable Star Observers (AAVSO) using the 20-inch Telescope Altiplano de Granada (TAGRA) in Spain in Bessel and Sloan/SDSS bands. In the light curves assembled from this photometry (Figure~\ref{fig:lc_full}), the typical optical photometric error bars are smaller than the symbol sizes. For example, typical ZTF uncertainties were $\lesssim$0.05~mag in $g$ and $\lesssim$0.02~mag in $r$. In the infrared, PGIR $J$-band uncertainties were $\sim$0.25~mag during the outburst, but they were significantly larger when the source was near the detection threshold before brightening. In contrast, the median VVV $JHK_s$ uncertainties were each $<$0.03~mag, providing good constraints on the pre-outburst brightness. Uncertainties on the NEOWISE photometry were $\sim$0.08~mag and $\sim$0.05~mag for $W1$ and $W2$, respectively.

Spectroscopic follow-up in the optical and near-infrared was obtained whilst the source was near maximum brightness, during the decay, and after its return to quiescence. These observations (Table~\ref{tab:obslog}) were made with SpeX \citep{Rayner2003} from NASA's Infrared Telescope Facility (IRTF), the Low Resolution Imaging Spectrometer \citep[LRIS;][]{Oke1995} on Keck~I, the Kast spectrograph \citep{miller1993} on Lick Observatory's 3-meter Shane Telescope, and the Triple Spectrograph \citep[TripleSpec;][]{Herter2008} on Palomar Observatory's 200-inch Hale Telescope. SpeX and TripleSpec data were reduced with Spextools \citep{Cushing2004}, LRIS with custom Interactive Data Language (IDL) scripts, and Kast with PypeIt v1.13.0 \citep[][]{2020JOSS....5.2308P,pypeit1.13.0}.

The LRIS and SpeX spectra, taken three days apart, overlap in the $Z$ and $Y$ bands. The shapes of the observed continua are consistent in this region. However, there is a factor of $\sim$1.6 discrepancy between the recorded fluxes in the LRIS (higher) and SpeX (lower) spectra. This shift could represent source variability but could alternatively be attributed to light lost from the slit or uncertainty in absolute flux calibration. However, there is no extended emission around the source that could cause this discrepancy when the slit sizes differ. Given that this offset can cause difficulties in analysis, we scaled the spectra to match in the overlap region and used the closest-in-time PGIR measurement ($J=12$~mag measured 27 days earlier) for absolute calibration. We apply scale factors of 0.9 to LRIS and 1.44 to SpeX to bring them into agreement. In the LRIS spectrum, the $\uplambda$8662 Ca\,{\sc ii} line is affected by a bad pixel, so we measured the Ca triplet lines from SpeX instead. 

\section{The Outburst}\label{sec:outburst}

The Gaia alert, Gaia22dtk, was based on a 5 Sep 2022 Gaia $G$-band measurement showing V1741~Sgr brighter than its historical level by 2.7~mag \citep[][]{Hodgkin2021,2022TNSTR2604....1H}.\footnote{\url{http://gsaweb.ast.cam.ac.uk/alerts/alert/Gaia22dtk/}} The rise also triggered multiple alerts from the ZTF (ZTF18abfogsw) and Gattini (PGIR22agkate) surveys. 

\begin{figure}
    \includegraphics[width=0.48\textwidth]{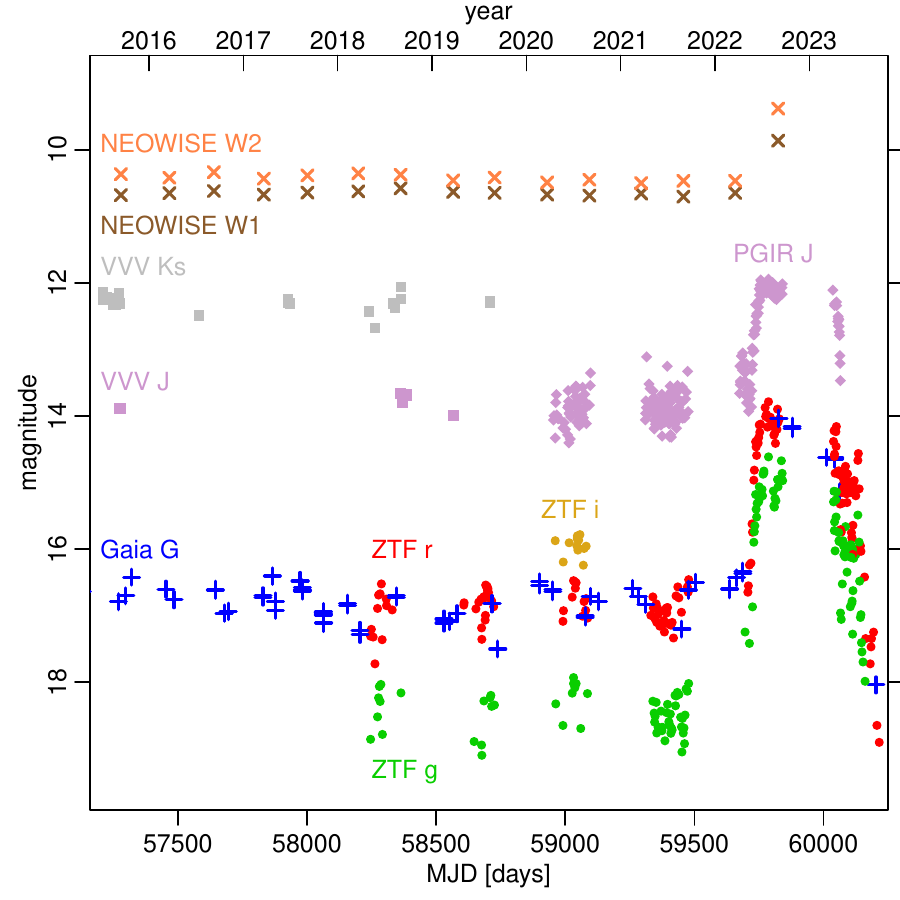}
    \caption{Optical and infrared light curves for V1741~Sgr over an $\sim$8~year period from 2015 to mid-2023. The light curves come from Gaia, ZTF, PGIR, VVV, and NEOWISE, and for $\mathrm{MJD} > 60000$ the ZTF photometry have been supplemented by AAVSO measurements.  Most error bars are smaller than the symbols, so they are not depicted.
    }
    \label{fig:lc_full}
\end{figure}

\begin{figure}
    \includegraphics[width=0.48\textwidth]{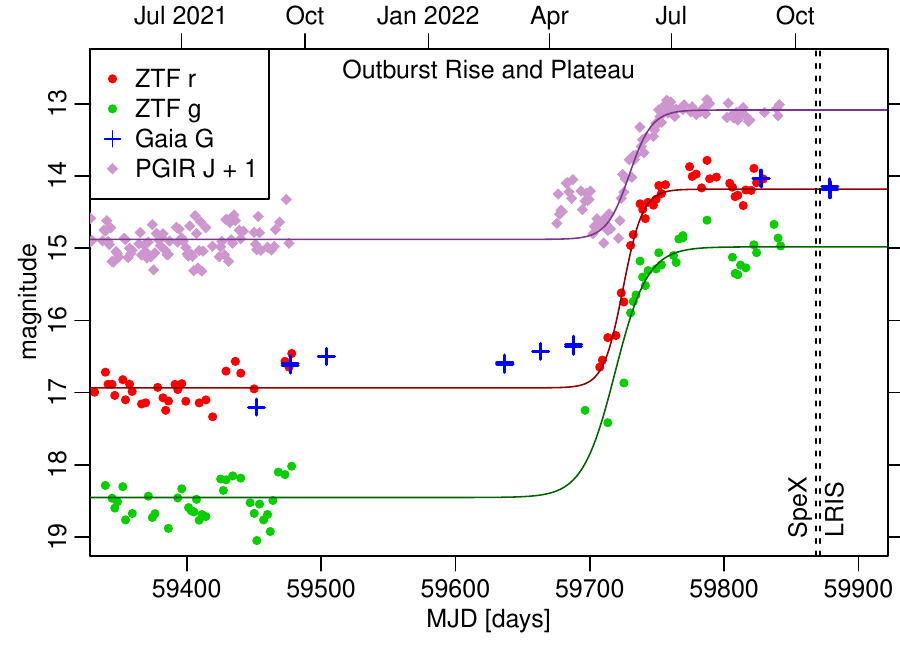}
    \caption{Zoom-in on the outburst rise as seen in the ZTF $g$ and $r$-bands, the Gaia $G$-band, and the PGIR $J$-band. Logistic functions are fit to the ZTF data (red and green curves) and the combined VVV (not shown) plus PGIR data (purple curve). The vertical dashed lines indicate the dates of the first two spectroscopic observations.
    }
    \label{fig:lc_fit}
\end{figure}

The outburst's rise occurred during early-to-mid 2022, and was detected in the Gaia $G$-band, ZTF's $g$ and $r$-bands, the PGIR $J$-band, and WISE's $W1$ and $W2$ bands (Figures~\ref{fig:lc_full} and \ref{fig:lc_fit}). In the Gaia light curve, the source brightened from a historical level of $G \sim 16.8$~mag to its alert magnitude of $G= 14.0$~mag, the brightest Gaia measurement to date. The last pre-outburst Gaia measurement was made on 18 April 2022, with $G=16.3$~mag, which is close to the historical mean, indicating that most of the rise occurred between mid-April and early-September of that year. ZTF recorded the source brightening from $g\sim18.4$ to 15.0~mag ($\Delta g \sim 3.4$~mag) and $r\sim16.9$ to 14.2~mag ($\Delta r \sim 2.7$~mag), VVV+PGIR recorded the source brightening from $J \sim 13.8$ to 12.0~mag ($\Delta J \sim 1.8$~mag), and NEOWISE recorded the source brightening from $W1 \sim 10.6$ to 9.9~mag ($\Delta W1\sim0.8$~mag) and $W2\sim10.4$ to 9.4~mag ($\Delta W2\sim1.0$~mag). The brightening at visible wavelengths exceeds the factor-of-ten flux increase typically used as a threshold to classify events as outbursts \citep{Fischer2023}. Larger amplitudes at shorter wavelengths match expectations for YSO outbursts \citep{2022RNAAS...6....6H}.

PGIR's observability windows start slightly earlier than ZTF, providing finer detail about the initiation of the outburst (Figures~\ref{fig:lc_fit}). This light curve showed a temporary $\sim$1~mag brightening in April 2022, just before the start of the main rise. 

Once V1741 Sgr reached maximum brightness in mid-July 2022, the light curve plateaued, lasting until the source became unobservable from Palomar for several months during the winter season. There was a temporary $\sim$0.5~mag dip in August of 2022. 
 
\begin{figure*}
    \includegraphics[width=1\textwidth]{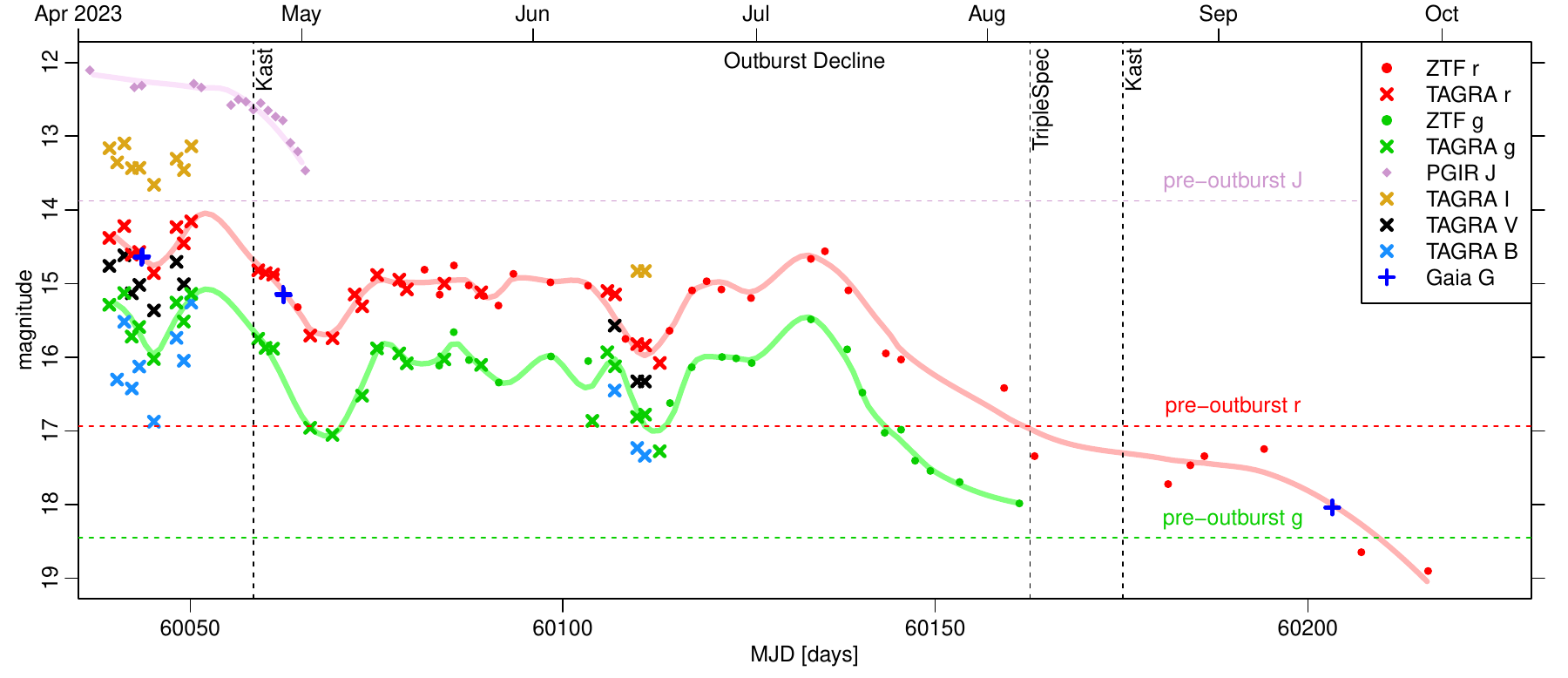}
    \caption{Multi-band light curves for V1741 Sgr during the Spring and Summer of 2023 while the source was dimming. Photometry include the PGIR $J$ band, the ZTF $g$ and $r$ bands, the Gaia $G$ band, and the TAGRA measurements in the Johnson-Cousins $B$, $V$, and $I_c$ bands and the Sloan/SDSS $g$ and $r$ bands. Smoothed curves (loess regression) for the $g$, $r$, and $J$ light curves are provided to guide the eye. Formal photometric uncertainties are smaller than the symbols, so error bars are not shown. Dates of spectroscopic observations are indicated by the vertical dashed lines, and the pre-outburst mean magnitudes are indicated by the horizontal dashed lines.}
    \label{fig:decay}
\end{figure*}

The decay of the outburst was captured during the spring and summer of 2023, when it faded from near-peak brightness to fainter than its pre-outburst state (Figure~\ref{fig:decay}). The first infrared and optical observations of 2023 came in April, with the source still in its plateau phase but showing some variability and a slight decline from the previous summer. However, by the end of that month, the decay had definitively begun. The next three months showed both dimming and rebrightening events, but the source finally reached its pre-outburst magnitude in August 2023 and dipped below this threshold in the subsequent months.  

The history of monitoring V1741~Sgr extends back to 2015 for Gaia and 2010 for VVV and (NEO)WISE, but no previous burst or outburst was detected.

\begin{figure}
    \includegraphics[width=0.48\textwidth]{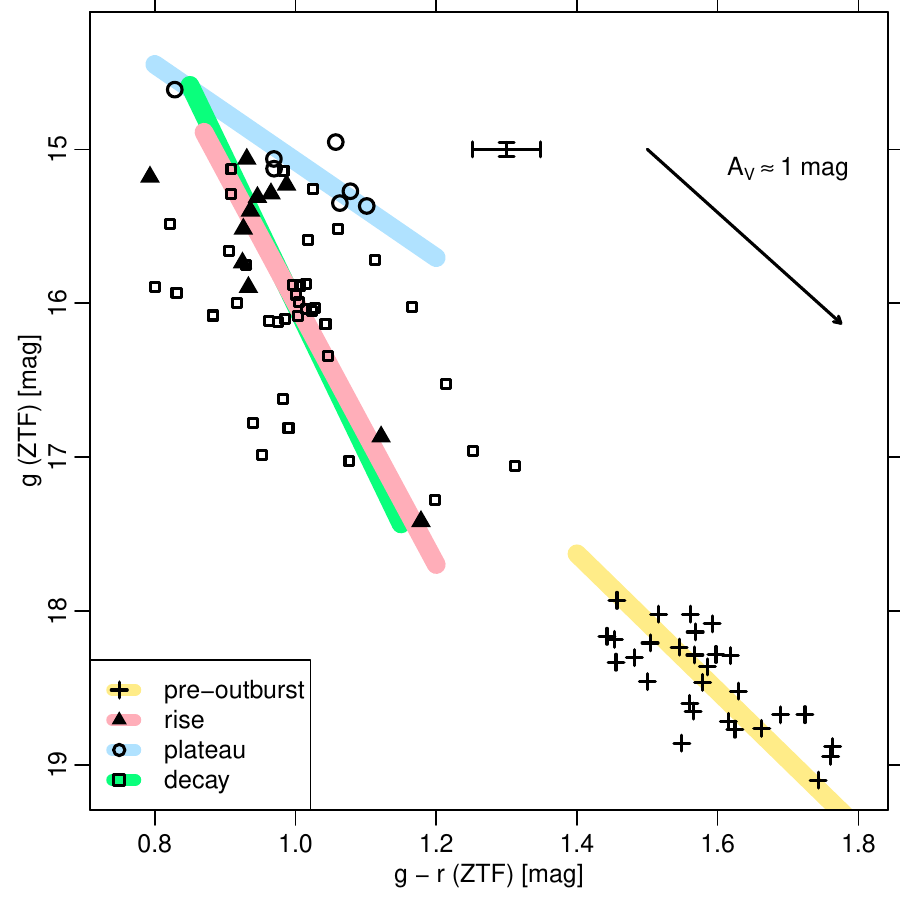}
    \caption{ZTF colour-magnitude diagram for the light curve's pre-outburst, rise, plateau, and decay phases. Typical error bars and an $A_V\approx1$~mag reddening vector bar are shown. The thick coloured lines show the orthogonal distance regressions for each phase of the light curve. 
    }
    \label{fig:color}
\end{figure}

\subsection{Modeling the Rise}\label{sec:sigmoid}

ZTF and PGIR sampled the light curve at multiple epochs along the entirety of the rise, revealing sigmoid-like behaviour (Figures~\ref{fig:lc_fit}). The logistic curve, 
\begin{equation}
m(t) = m_0 - \Delta m\,\left(1 + \exp\left((t-t_0)/\tau\right)\right)^{-1},
\end{equation}
provides a convenient functional form to model the observed rises \citep[e.g.,][]{Hillenbrand2018}. Here, $m_0$ is the quiescent magnitude, $\Delta m$ is the change in magnitude, $t_0$ is the time of the midpoint of the rise in MJD,
 and $\tau$ sets the rise timescale. In the $r$-band, which is the best sampled, the middle of the rise occurs at $t_0 = 59724.8\pm0.9$~days (25 May 2022), and the rise has a timescale of $\tau=7.3\pm0.8$~days. This timescale indicates that $\sim$90\% of the rise occurred in 2 months, with a maximum rise rate of $\mathrm{d}r/\mathrm{d}t = -0.094$~mag~day$^{-1}$ for the fitted curve. For the $g$-band, $t_0 = 59718.9\pm1.8$~days and $\tau = 12.1\pm1.5$~days. To fit the $J$ band, we used previous VVV measurements to constrain the pre-outburst magnitude and PGIR measurements for the outburst, finding $t_0 = 59731.2\pm1.5$~days and $\tau=8.4\pm1.4$~days. 

These fits indicate that the $g$, $r$, and $J$ light curves were not synchronised but showed wavelength-dependent lags in when they reached their midpoints. The $g$-band light curve reached its midpoint first, followed by the $r$-band light curve with a delay of 6 days, and finally, the $J$-band light curve 6 days after that. These delays can be seen with visual inspection of Figure~\ref{fig:lc_fit}, are statistically significant, and are similar to the $\tau$ brightening timescales. Each of the logistic-function fits is baselined to the long-term low-state of the light curves from ZTF or VVV, so these differences are not produced by slight differences in the optical and near-infrared observing seasons.

In the NEOWISE light curve, sampled every six months, only the last epoch (MJD 59824) showed significant brightening. This timing is consistent with the optical/near-infrared outburst (Figure~\ref{fig:lc_full}) and indicates that V1741~Sgr did not brighten in the mid-infrared pre-outburst.

\subsection{Outburst Decay}\label{sec:decline}

The shape of the $g$ and $r$-band light curves during the decay
do not mirror the rise, but instead, they are choppier with multiple dimming and rebrightening events amidst a net decline (Figure~\ref{fig:decay}). Consequently, the sigmoid-like models from Section~\ref{sec:sigmoid} do not describe these curves well. Once dimming began, the time for the source to reach its pre-outburst flux was $\sim$110~days ($\sim$3.6~months), with a mean rate of $\mathrm{d}r / \mathrm{d}t =0.025$~mag~day$^{-1}$. This implies that the dimming phase of this outburst was slightly slower than the rising phase. 

Local minima in the decay light curve are found at MJD 60045, 60067, and 60111. During the first of these, the light curve faded to near its pre-outburst magnitude in the $J$ band. During a local maximum at MJD 60135, the source recovered much of its lost flux, before beginning a final rapid dimming. 

In the final measurements before our cut-off date, the source became fainter than the pre-outburst level by $\Delta r\approx2$~mag. This extra dimming was detected in one Gaia measurement and two $r$-band ZTF measurements. No previous dip of this depth had been recorded in monitoring of this source, suggesting that the dimming could have been triggered by the outburst.

\subsection{Optical Colour Changes}\label{sec:color}

Overall, bluer $g-r$ colours are correlated with brighter $g$-band magnitudes (Figure~\ref{fig:color}). This behaviour is common for variable YSOs, where it can be explained by changes in extinction or accretion flows \citep[e.g.,][]{2003A&A...409..169B,2018AJ....155...99W,2022AJ....163..263H}. The \citet{Cardelli89} extinction law with $R_V=3.1$ yields the reddening relation\footnote{As this relation depends slightly on the source \citep[e.g.,][]{2009MNRAS.392..497S}, we assumed a K-type stellar spectrum when calculating approximate reddening coefficients for the ZTF filters.} $E(g-r) \approx 0.275 A_V$ and $A_g \approx 1.13 A_V$.

To analyse the colour variability of V1741 Sgr, we subdivided the light curves into phases, including pre-outburst (MJD~58200--59600), rise (MJD~59600--59777), plateau (MJD~59777--59850), and decay (MJD~60030--60170). To calculate colours, we matched ZTF $g$ and $r$-band observations made within 12 hours of each other.

Regression lines are fit to the $g$ vs.\ $g-r$ data points in each phase of the light curve. We used the orthogonal distance regression \citep{golub1980analysis}, implemented in the CRAN package {\it pracma} \citep{pracma}. This regression algorithm is well-suited for analysing these diagrams because it treats both colour and magnitude in a symmetric manner \citep{2015AJ....150..118P,2022AJ....163..263H}. 

Whilst in its pre-outburst state, V1741~Sgr exhibited significant variability, with standard deviations of $0.3$~mag in $g$ and 0.25~mag in $r$. The slope for this phase, $\Delta g/\Delta(g-r) = 4.5\pm0.8$ (yellow line), is consistent with the reddening vector ($\mathrm{slope} = 4.1$), suggesting that this variability may be attributed to extinction by circumstellar material.

At the start of the outburst, the source initially became bluer by $\Delta(g-r) = 0.4$~mag, but then, during the rise phase, the source's brightness increased dramatically, with relatively little colour change. For this phase, the slope is $8.5\pm1.2$ (red line), significantly steeper than the reddening vector, implying that this brightness change is driven by something other than extinction. The initial blueward colour shift is connected to the $\sim$6-day lag between the $g$ and $r$-band rises.

During the plateau, the slope of $3.1\pm1.0$ (blue line) is again consistent with variability produced by changes in extinction. Thus, the temporary dip in mid-August may have been an extinction-related event. 

The slope during the decline is $9.5\pm3.0$ (green line), similar to that of the rise. However, there is considerably more scatter in the $g$ vs.\ $g-r$ diagram than during the earlier stages of the light curve. The width of the distribution is statistically significant, which may suggest that multiple phenomena contribute to the colour changes, possibly both extinction and accretion changes. 

\begin{figure*}
    \includegraphics[width=1\textwidth]{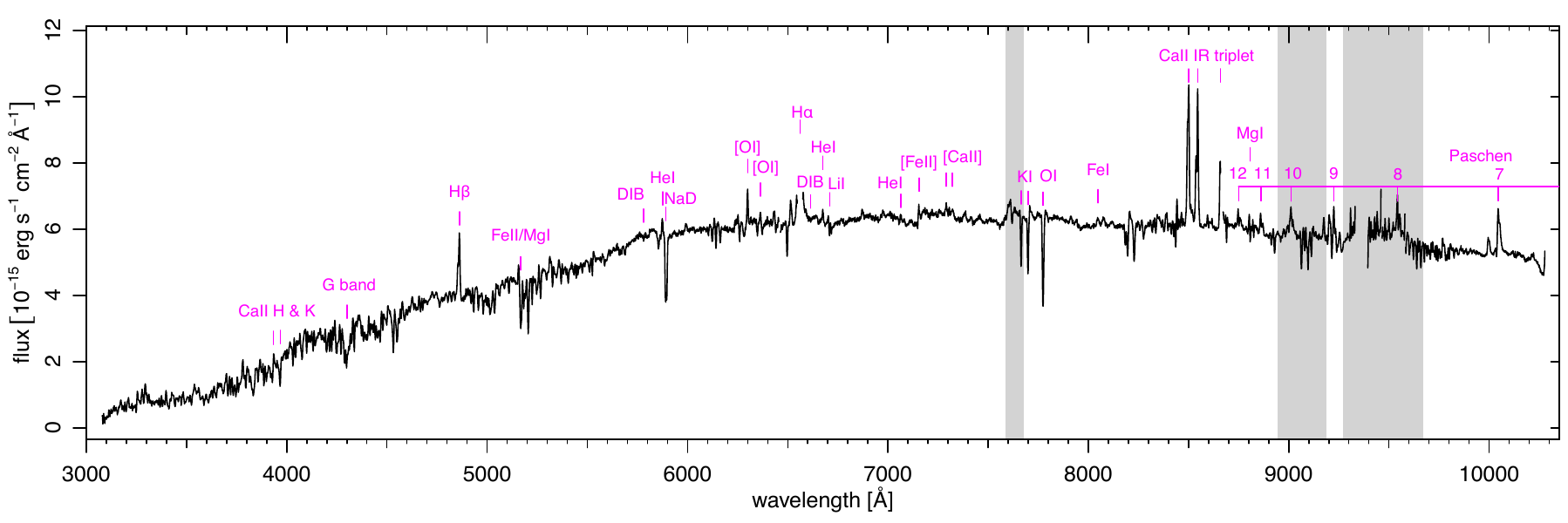}
    \includegraphics[width=1\textwidth]{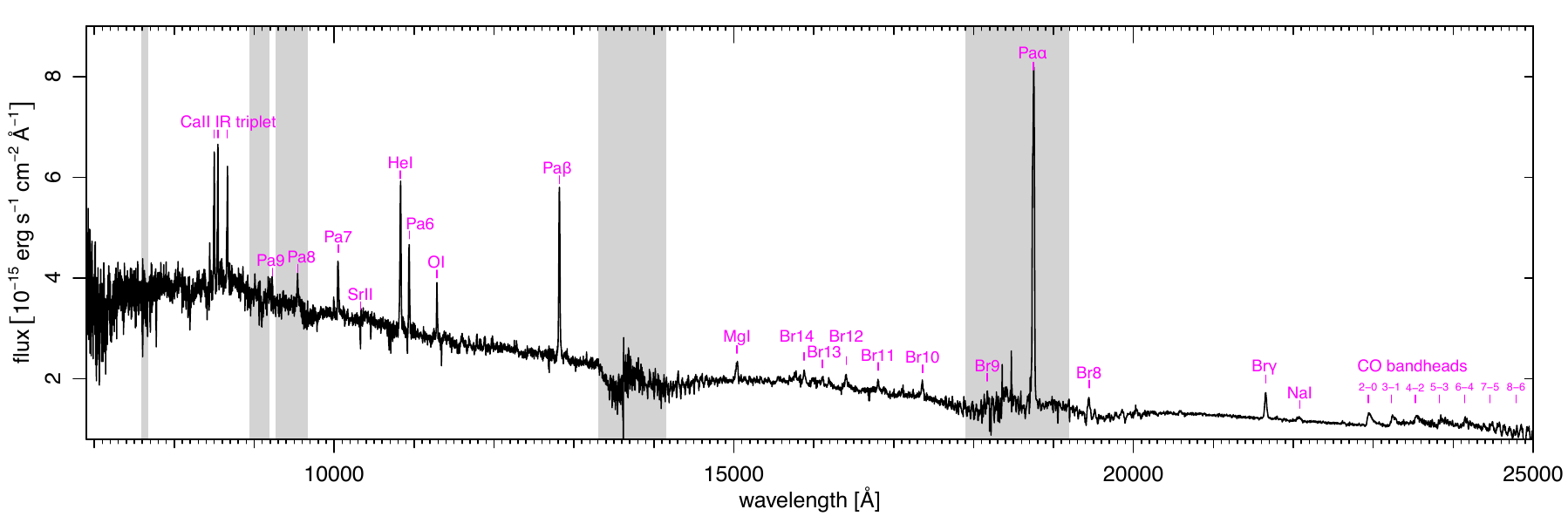}
    \caption{Spectra of V1741~Sgr during the outburst plateau from Keck/LRIS (top) and IRTF/SpeX (bottom). Prominent features are indicated. These spectra are shown as they appear in the original data, without the scaling described in Section~\ref{sec:data}. In both panels, spectroscopic regions affected by strong atmospheric absorption are shaded grey
    }
    \label{fig:spec}
\end{figure*}

\section{Spectroscopy of the Outburst Peak}\label{sec:spec}

At the outburst peak, the LRIS and SpeX spectra reveal a source with both emission and absorption features (Figure~\ref{fig:spec}). The strongest emission lines include the hydrogen Balmer, Paschen, and Brackett series, the Ca\,{\sc ii} infrared triplet (the Ca\,{\sc ii} H and K lines are not unambiguously detected), and He\,{\sc i} lines (Table~\ref{tab:ew}). Other emission lines include Fe\,{\sc i} and Mg\,{\sc i}, the Na\,{\sc i} doublet at $\uplambda\uplambda$2.206/2.209~$\upmu$m, and forbidden [O\,{\sc i}], [N\,{\sc ii}], [Ca\,{\sc ii}], and [Fe\,{\sc ii}] lines. The CO overtone bands are also detected in emission. In the optical, the spectrum is fairly red as a consequence of extinction (Section~\ref{sec:extinction}). However, in the near-infrared, the source has a blue continuum.

\begin{table}
	\centering
	\caption{Emission Lines (Outburst State)}
	\label{tab:ew}
	\begin{tabular}{lrrrr} 
		\hline
		\multicolumn{1}{c}{Line} & 
		\multicolumn{1}{c}{$\lambda$} & 
		\multicolumn{1}{c}{$W_\lambda$} &
		\multicolumn{1}{c}{$f_\mathrm{obs}$} & 
		\multicolumn{1}{c}{$L_\mathrm{line}$} \\
           & 
               \multicolumn{1}{c}{ \AA} & 
                \multicolumn{1}{c}{\AA}  & 
                \multicolumn{1}{c}{erg\,s$^{-1}$\,cm$^{-2}$} & 
                \multicolumn{1}{c}{erg\,s$^{-1}$} \\
                &  &   & 
                \multicolumn{1}{c}{$10^{-15}$} & 
                \multicolumn{1}{c}{$10^{30}$} \\
  \multicolumn{1}{c}{(1)} & 
  \multicolumn{1}{c}{(2)} & 
  \multicolumn{1}{c}{(3)} & 
  \multicolumn{1}{c}{(4)} & 
  \multicolumn{1}{c}{(5)} \\
		\hline
H$\beta$ & 4861 & $-5.01\pm0.25$ & $17.9\pm0.9$ & $15\pm6$ \\\relax
He\,{\sc i} & 5876 & $-0.64\pm0.03$ & $3.3\pm0.2$ & $2\pm1$ \\\relax
[O\,{\sc i}] & 6300 & $-0.96\pm0.01$ & $5.3\pm0.1$ & $3\pm1$ \\\relax
O\,{\sc i} & 8446 & $-1.73\pm0.19$ & $9.4\pm1.0$ & $4\pm1$ \\\relax
Ca\,{\sc ii} & 8498 & $-7.82\pm0.27$ & $43.3\pm1.5$ & $16\pm3$ \\\relax
Ca\,{\sc ii} & 8542 & $-8.03\pm0.26$ & $44.7\pm1.5$ & $17\pm3$ \\\relax
Ca\,{\sc ii} & 8662 & $-6.88\pm0.25$ & $38.8\pm1.4$ & $14\pm2$ \\\relax
Pa9 & 9230 & $-2.03\pm0.09$ & $9.9\pm0.5$ & $3\pm1$ \\\relax
Pa8 & 9546 & $-2.16\pm0.19$ & $10.9\pm0.9$ & $4\pm1$ \\\relax
Pa7 & 10,049 & $-4.81\pm0.12$ & $22.8\pm0.6$ & $7\pm1$ \\\relax
He\,{\sc i} & 10,830 & $-16.78\pm0.29$ & $72.2\pm1.2$ & $22\pm3$ \\\relax
Pa6 & 10,938 & $-7.91\pm0.24$ & $33.9\pm1.0$ & $10\pm1$ \\\relax
O\,{\sc i} & 11,287 & $-4.46\pm0.15$ & $18.1\pm0.6$ & $5\pm1$ \\\relax
Pa$\beta$ & 12,818 & $-24.60\pm0.29$ & $86.9\pm1.0$ & $23\pm2$ \\\relax
Mg\,{\sc i} & 15,040 & $-6.23\pm0.17$ & $17.6\pm0.5$ & $4.4\pm0.3$ \\\relax
Pa$\alpha$ & 18,751 & $-134.02\pm0.96$ & $284.0\pm2.0$ & $65\pm3$ \\\relax
Br$\gamma$ & 21,661 & $-13.10\pm0.24$ & $22.9\pm0.4$ & $5.0\pm0.2$ \\\relax
Na\,{\sc i} & 22,080 & $-3.48\pm0.21$ & $5.8\pm0.4$ & $1.3\pm0.1$ \\
  \hline
\end{tabular}
\flushleft{Columns 1--2: Spectral line identity. Below 20,000~\AA, wavelengths are given in air and, above, in vacuum. Column 3: Equivalent width. Column~4: Observed line flux. Column 5: Extinction-corrected line luminosity assuming $A_V = 1.4\pm0.4$~mag and $\varpi = 0.795$~mas.}
\end{table}

The strong emission lines, and especially the CO emission, indicate that the outburst from V1741~Sgr best fits into the EX Lup category \citep{Fischer2023}. The Li\,{\sc i} $\uplambda$6707 absorption line is measured\footnote{
To correct for blending with the nearby iron line, we use the relation from \citet{2023MNRAS.tmp.1266J} to estimate the contribution from $\uplambda$6707.44 Fe\,{\sc i} for a stellar atmosphere with $T_\mathrm{eff} \approx 4750~K$ (Section~\ref{sec:optical_spectrum}). The expected contribution of $0.02$~\AA\ is much smaller than the observed equivalent width.
}
 in the LRIS spectrum with an equivalent width of $W_\lambda = 0.28\pm0.03$~\AA, providing further evidence that V1741~Sgr is a young star. 

Within the EX Lup category, there is significant diversity in spectroscopic features, both in terms of which features are present and the strengths of these features. To illuminate similarities and differences within this class, in Figures~\ref{fig:baII_srii}--\ref{fig:jhk_zoom}, we compare our spectra of V1741~Sgr to two other outbursting YSOs, the Class~I source Gaia~19ajj with a 5.5~mag rise over 3 years \citep{Hillenbrand2019} and the intermediate-mass star LkH$\alpha$~225~S with a 7~mag rise over 15 years \citep{Magakian2019,Hillenbrand2022}. 

\subsection{Optical/Near-Infrared LRIS Spectrum}\label{sec:optical_spectrum}

In the optical spectrum (Figure~\ref{fig:spec}, top), absorption features resemble those of a K or G-type star, including the presence of the Fraunhofer G band \citep{GrayCorbally2009}. We modelled the normalised spectrum between 4000~\AA\ and 4500~\AA, masking out the H$\gamma$ and H$\delta$ lines, using the BOSZ \citep{Bohlin2017} stellar atmospheres plus a constant component to approximate the effect of veiling on spectral lines and assuming Solar metallicity. A 4750~K stellar atmosphere with a veiling component $\sim$1.4 times the continuum level provided the best fit. However, the degeneracy between temperature and veiling, with higher veiling yielding lower temperatures, means these parameters are not well constrained. 

Several lines present deeper absorption than a normal stellar photosphere, including the Mg\,{\sc i}\,b triplet, the Na\,{\sc i}\,D doublet, the $\uplambda$7665 and $\uplambda$7699 K\,{\sc i} lines, and the $\uplambda\uplambda\uplambda$7774 O\,{\sc i} triplet. These lines are frequently associated with winds in YSO spectra \citep[e.g.,][]{Covey2011,Hillenbrand2018,Hillenbrand2022}; however, the resolution of our V1741~Sgr spectrum is too low to discern the kinematics. 

\begin{figure}
    \includegraphics[width=0.48\textwidth]{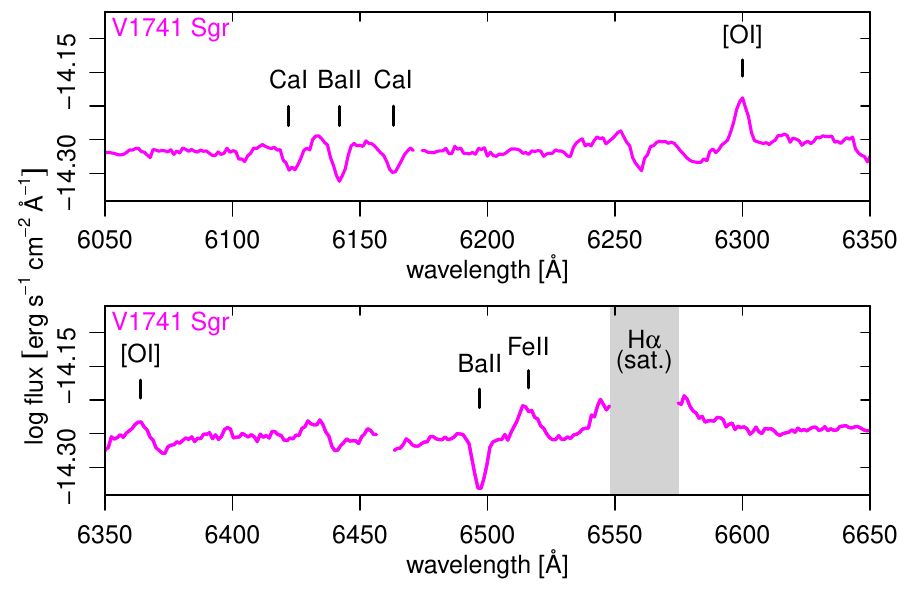}\\
    \includegraphics[width=0.48\textwidth]{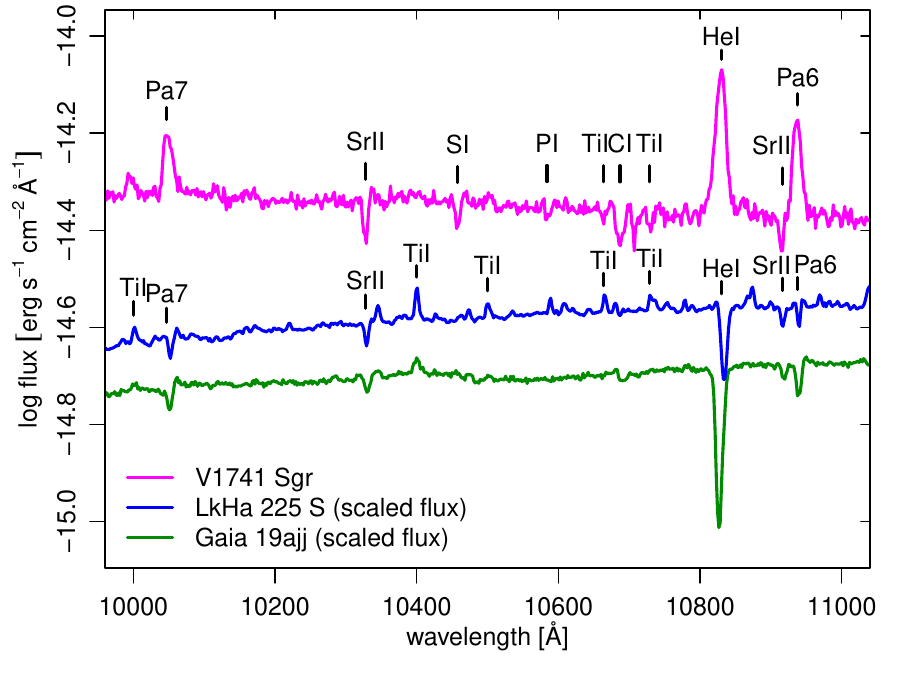}
    \caption{Detailed views of the outburst spectra in the $r$-band (top) and $Y$-band (bottom). In the bottom panel, SpeX spectra for Gaia 19ajj and LkH$\alpha$ 225 S are included for comparison. 
    }
    \label{fig:baII_srii}
\end{figure}

\begin{figure*}
    \includegraphics[width=1\textwidth]{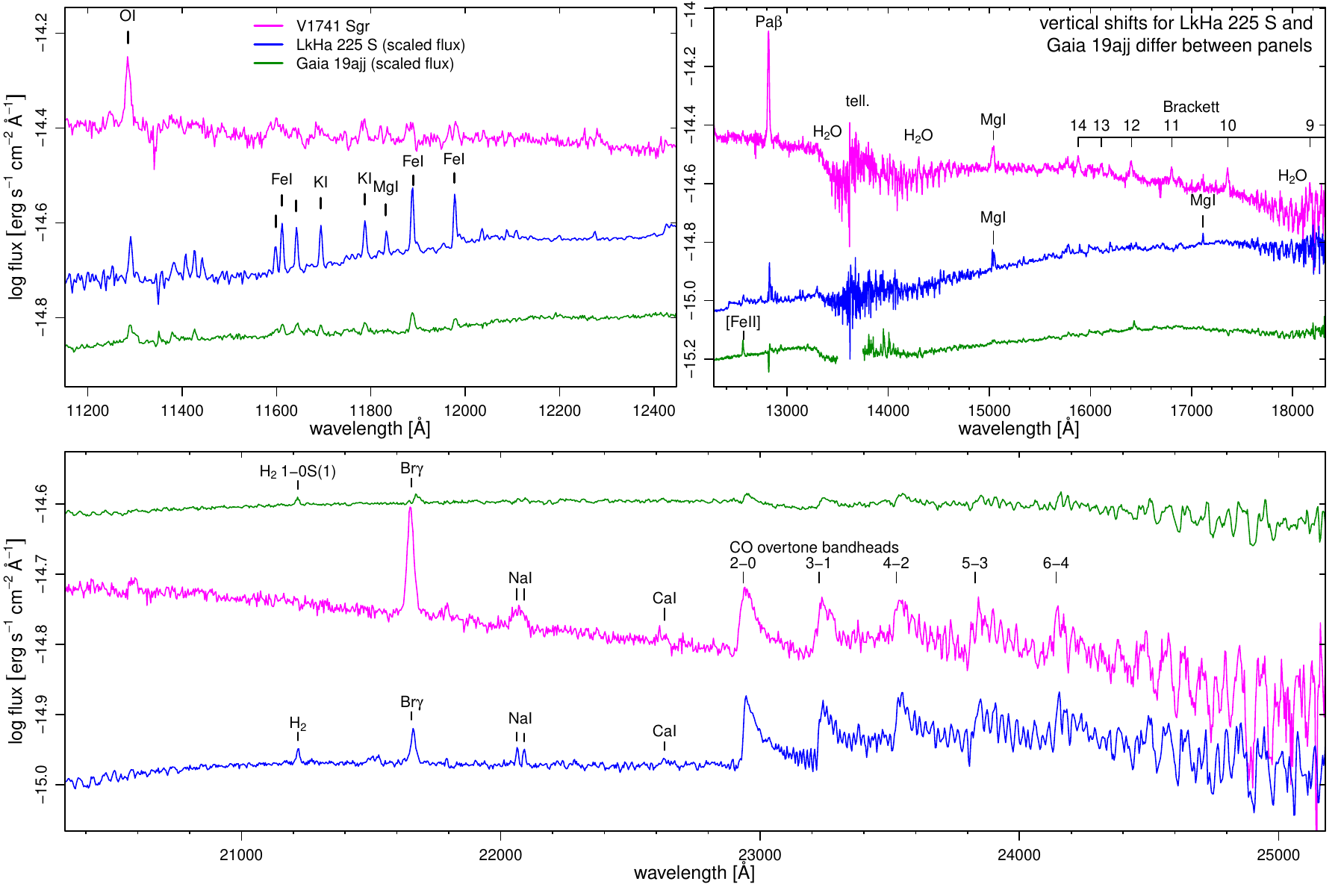}
    \caption{Detailed views of the outburst SpeX spectra for V1741~Sgr (magenta), LkH$\alpha$~225~S (blue), and Gaia~19ajj (green) in spectral regions around the $J$ (top left), $H$ (top right), and $K$ (bottom) bands. 
    }
    \label{fig:jhk_zoom}
\end{figure*}

The spectrum of V1741~Sgr shows Ba\,{\sc ii} absorption lines at $\uplambda$6142 and $\uplambda$6497 (Figure~\ref{fig:baII_srii}, top). These are signatures of low-gravity atmospheres, including the atmospheres of circumstellar discs, and are commonly observed in FU Ori-type stars \citep[e.g.,][]{Reipurth1990,1997A&AS..126...91T,2019ApJ...874...82H}.   

\subsection{Near-Infrared SpeX Spectrum}\label{sec:spex}

In the $Y$-band (Figure~\ref{fig:baII_srii}, bottom), Sr\,{\sc ii} absorption lines provide further indication of low gravity. For F--M stars, the Sr\,{\sc ii} lines at $\uplambda$1.0328~$\upmu$m and  $\uplambda$1.0916~$\upmu$m are strongest in supergiants, moderate in giants, and weak in dwarfs \citep{Sharon2010,Hillenbrand2022}. These Sr\,{\sc ii} absorption lines are detected in V1741~Sgr, Gaia~19ajj, and LkH$\alpha$~225~S, but they are strongest in V1741~Sgr. 

The $Y$-band region of  V1741 Sgr's spectrum exhibits atomic absorption lines with low-to-moderate excitation potentials,
including the Sr\,{\sc ii} lines discussed above ($E_\mathrm{lower}\sim1.8$~eV), Ti\,{\sc i} ($E_\mathrm{lower}\sim0.8$~eV), and C\,{\sc i}, S\,{\sc i}, and P\,{\sc i} ($E_\mathrm{lower}\sim7$~eV). This region of the spectrum looks quite different for Gaia~19ajj or LkH$\alpha$~225~S, where the Ti\,{\sc i} lines are seen in emission and the C\,{\sc i}, S\,{\sc i}, and P\,{\sc i} lines are not detected. The higher excitation C\,{\sc i}, S\,{\sc i}, and P\,{\sc i} lines require temperatures of $>$4700~K to form in stellar atmospheres \citep{Sharon2010}. For example, in their sample of standard stars, C\,{\sc i}  $\uplambda$1.0688~$\upmu$m absorption is only detected from supergiants with $T_\mathrm{eff}\gtrsim4700$~K, giants with $T_\mathrm{eff}\gtrsim5100$~K, or dwarfs with $T_\mathrm{eff}\gtrsim5800$~K. 

In the $J$ band (Figure~\ref{fig:jhk_zoom}, top left), outbursting YSOs often exhibit numerous emission lines between 1.150~$\upmu$m and 1.210~$\upmu$m from Fe\,{\sc i}, Mg\,{\sc i}, Si\,{\sc i}, and K\,{\sc i} \citep[e.g.,][]{Kospal2011,Hodapp2019}. These can be seen in the spectrum of LkH$\alpha$~225~S, or, more weakly, in the spectrum of Gaia~19ajj \citep{Hillenbrand2022}. However, these lines are much weaker in the spectrum of V1741~Sgr. In contrast, the nearby $\uplambda$1.1287~$\upmu$m O\,{\sc i} line is relatively stronger in V1741~Sgr than in the other two spectra. 

A sharp down-turn in the spectrum of V1741~Sgr around 1.34~$\upmu$m indicates the onset of water absorption, and absorption features can also be seen on both sides of the $H$-band spectrum (Figure~\ref{fig:jhk_zoom}, top right).  Similar absorption features were detected in the Gaia~19ajj and LkH$\alpha$~225~S spectra \citep{Hillenbrand2019,Hillenbrand2022}, but the features are slightly weaker in those stars. Furthermore, the water absorption in each of these three stars is much weaker than the water absorption frequently seen in FU Ori stars, where the effect of low gravity can give the $H$-band a characteristic triangular shape \citep[e.g.,][]{Connelley2018}. Water absorption is typically found in the spectra of M-type stars, becoming prominent at lower temperatures than molecular absorption from TiO and VO \citep{GrayCorbally2009}. However, no TiO or VO absorption was detected in the LRIS or SpeX V1741~Sgr spectra.

Spectra of EX Lup-type stars frequently exhibit H$_2$ emission associated with shocked gas in outflows \citep[e.g.,][]{2020MNRAS.492..294G,2022ApJ...929..129G,2023ApJ...945L...7K}. However, no H$_2$ lines were detected in our spectrum of V1741~Sgr (Figure~\ref{fig:spec}, bottom), with an upper limit of $<\!\!4\times10^{-16}$~erg~s$^{-1}$~cm$^{-2}$ on the strongest of these lines at  $\uplambda$2.12~$\upmu$m. In contrast, the  $\uplambda$2.12~$\upmu$m line was detected in the spectra of both LkH$\alpha$~225~S and Gaia~19ajj, even while these stars were in outburst. 

The Na\,{\sc I} doublet at  $\uplambda\uplambda$2.206/2.209~$\upmu$m is detected as a blended emission line with an equivalent width of $-3.45$~\AA. It is stronger in V1741 Sgr than in either Gaia 19ajj (not detected) or in LkH$\alpha$ 225 S (weakly detected), but similar to the strength of the Na\,{\sc I} lines initially detected from EX Lup \citep[$-2.45$~\AA;][]{Kospal2011}.

In V1741~Sgr, the CO overtone bands are strong emission features (Figure~\ref{fig:jhk_zoom}, bottom), with a similar strength relative to the continuum as LkH$\alpha$~225~S and much stronger than for Gaia 19ajj (Figure~\ref{fig:jhk_zoom}, bottom). CO in emission is a common trait of EX Lup-type sources, whereas CO absorption is required in the definition of FU Ori-type stars \citep{Connelley2018}.
The strengths and profile shapes of the CO bands here resemble the outburst CO spectrum of EX Lup \citep[][their Figure~7]{Kospal2011}, which was found to be well-described by CO emission from a 2500 K slab model. 

\subsection{Diffuse Interstellar Bands and Extinction}\label{sec:extinction}

The extinctions of outbursting stars are challenging to measure owing to the unknown spectrum of the underlying sources. However, diffuse interstellar bands (DIBs) can serve as extinction estimators that do not require knowledge of spectral shape. In the LRIS spectrum, the DIBs at $\uplambda$5780 and $\uplambda$6614~\AA\ are detected with equivalent widths of 0.208$\pm$0.024~\AA\ and 0.072$\pm$0.018~\AA, respectively. \citet{2022ApJ...940..156C} note that DIB absorption may be contaminated by Fe\,{\sc i}, Cr\,{\sc i}, Si\,{\sc i}, and Sc\,{\sc ii} lines in stars with spectral types later than G0. In their sample of outbursting YSOs, they calculate that contamination may account for $\sim$0--50\% of the absorption in these features. Thus, the extinction values we calculate from these DIBs should be considered upper limits. 

The relations between DIB equivalent widths and extinctions \citep[][their Equations 4 and 5]{2022ApJ...940..156C} predict $E(B-V) = 0.45\pm0.12$ from the $\uplambda$5780 DIB or $E(B-V) = 0.35\pm0.11$ from the $\uplambda$6614 DIB. Assuming an extinction law with $R_V=3.1$, these correspond to $A_V\sim1.4$~mag or $1.1$~mag, respectively. 

Along the line of sight toward V1741 Sgr, the extinction estimated by the Bayestar \citep{Green2019} map increases gradually up to a distance of $\sim$1100-1200~pc, where extinction jumps by 0.6~mag from $A_V=0.6$~mag to $A_V=1.2$~mag, followed by another gradual increase at greater distances.\footnote{
Bayestar colours were converted to $A_V$ following the recommended extinction law from \citet{Green2019}.
}
This suggests the presence of a whispy local cloud in the vicinity of V1741~Sgr. Together with the DIB measurements, these results constrain V1741 Sgr's extinction during outburst to $A_V=0.6$--$1.4$~mag. If V1741~Sgr is partially obscured by its local cloud, it likely lies near the upper end of this range.

\begin{figure}
    \includegraphics[width=0.5\textwidth]{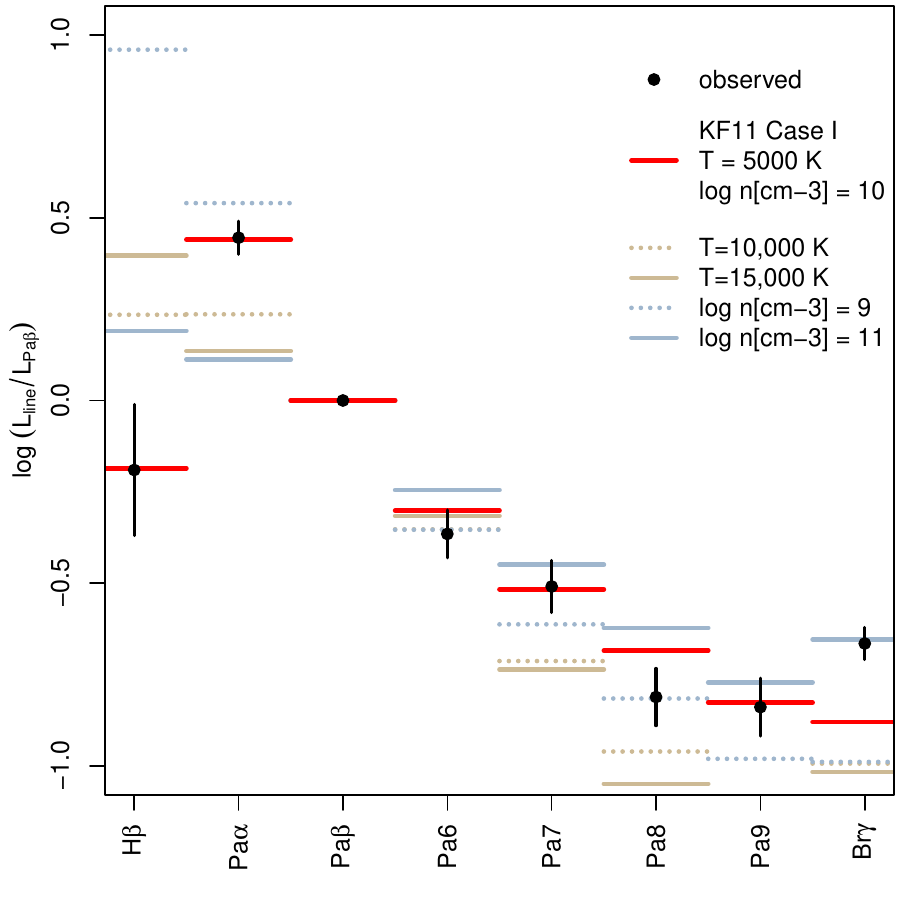}
    \caption{Ratio of the luminosities of various H lines to Pa$\beta$. The logarithmic line ratio is the ordinate, and the identity of the line in the numerator is the abscissa. The red lines indicate the ratios from the best-fitting model from \citet{2011MNRAS.411.2383K} Case~I. Other models with different temperatures or densities are included for comparison. 
    }
    \label{fig:line_ratios}
\end{figure}

\subsection{H\,{\sc i}, He\,{\sc i}, Ca\,{\sc ii}, and O\,{\sc i} Emission Lines}\label{sec:kf}

Hydrogen emission line ratios probe the conditions of the emitting gas. We compare emission lines to predictions for models of T Tauri stars with a photoionisation rate of $\gamma_{\mathrm{H}\,\textsc{i}} = 2\times10^{-4}$~s$^{-1}$  \citep{2011MNRAS.411.2383K}.\footnote{\url{https://www.stsci.edu/~wfischer/line_models.html}} For this analysis, we used H emission lines with adequate signal-to-noise and computed the ratio of the extinction-corrected line luminosities to the Pa$\beta$ line (Figure~\ref{fig:line_ratios}). The best fit was provided by a model with a temperature of $T=5000$~K and density of $\log n[\mathrm{cm}^{-3}] = 10$. This fit accurately reproduces the ratio of H$\beta$ to Paschen lines and the Paschen decrement. However, it underestimates the luminosity of the Br$\gamma$ line by $\sim$50\%. The good fit to the Balmer and Paschen lines implies that the assumed extinction of $A_V\lesssim1.4$~mag is reasonable. Modifying extinctions is insufficient to resolve the discrepancy for the Br$\gamma$ line since the Pa$\alpha$ and Br$\gamma$ lines are in the same spectral region and thus subject to similar extinctions. 

Figure~\ref{fig:line_ratios} also compares the results to several alternative models. Models with higher temperatures overestimate H$\beta$ and underestimate Pa$\alpha$ and Pa7--9. A model with ten times lower density overestimates the line strength of H$\beta$ by more than dex. Finally, a model with ten times higher density provides a slightly worse fit to H$\beta$ and the Paschen decrement than our best-fitting models but provides an improved fit to Br$\gamma$. Thus, it is possible that the discrepancy between the Br$\gamma$ and our best-fit model could be explained if the Br$\gamma$ emission was dominated by a slightly denser plasma. Fitting using the lower photoionization rate ($\gamma_{\mathrm{H}\,\textsc{i}} = 2\times10^{-6}$~s$^{-1}$) models from \citet{2011MNRAS.411.2383K} yielded similar results.

Our best-fitting density falls in the range found for other EX Lup-type stars observed in a high state, including XZ Tau, NY Ori, and PV Cep with $9 \lesssim \log n \lesssim 11$ \citep{2022ApJ...929..129G}. However, the best-fitting temperature for V1741~Sgr lies at the lower end of the temperature range $5000~\mathrm{K} \lesssim T \lesssim 12,500$~K.  

The Pa$\gamma$/He\,{\sc i}\,$\uplambda$10830 ratio for V1741 Sgr is $\sim$0.5, which implies that the Pa$\gamma$ line is moderately optically thick ($\tau_{\mathrm{Pa}\gamma} \sim 0.8$) according to the models of \citet[][their Figure~8]{2011MNRAS.411.2383K}. The He\,{\sc i}\,$\uplambda$5876/He\,{\sc i}\,$\uplambda$10830 ratio of $\sim$0.06, implies that the He\,{\sc i}~$\uplambda$5876 line is very optically thin and is lower than observed in the T Tauri stars examined by \citet[][their Figure~10]{2011MNRAS.411.2383K}. According to their models, these ratios constrain the temperature to 5000--15,000~K, consistent with the above values. 

The Ca\,{\sc ii} infrared triplet lines are observed strongly in emission, with no absorption component. Their line ratios (obtained from the SpeX spectrum) are close to unity, implying that they are optically thick, with a peak intensity pattern $I_\mathrm{\uplambda8542} >  I_\mathrm{\uplambda8498} > I_\mathrm{\uplambda8662}$ that is fairly typical for T Tauri stars \citep{Hamann1992,Azevedo2006}. In contrast, the Ca\,{\sc ii} H and K lines are not unambiguously detected despite sharing upper energy levels with the Ca\,{\sc ii} infrared triplet \citep{1943PASP...55..242M}. A comparable scenario was observed in the outburst spectrum of LkH$\alpha$ 225 S, where both emission and absorption P Cygni components were observed for the Ca II infrared triplet, but Ca\,{\sc ii} H and K were only detected in absorption \citep{Hillenbrand2022}. In the infrared, the Ca\,{\sc ii} triplet emission would be much stronger than the Ca\,{\sc ii} triplet absorption, so only the emission component is visible in V1741~Sgr's spectrum. Meanwhile, in the blue part of the spectrum, the absorption component is deeper, so it is only just filled by the emission component, leading to an absence of these lines. A similar scenario may explain the lack of Balmer lines blueward of H$\beta$.

The forbidden [Ca\,{\sc ii}] $\uplambda$7291 and $\uplambda$7324 transitions, whose upper state is the lower state of the permitted Ca\,{\sc ii} triplet \citep{1943PASP...55..242M}, are detected weakly in emission by LRIS. These lines are fairly rare in YSOs but observed during some outbursts, including Gaia~19ajj and LkH$\alpha$~225~S \citep{Hillenbrand2019,Hillenbrand2022}. 

The O\,{\sc i} $\uplambda$8446 and $\uplambda$11287 lines are sensitive to Ly$\beta$ irradiation, which can excite ground-state O\,{\sc i} to the 3d $^3$D state, which may then de-excite via sequential emissison of $\uplambda$8446, $\uplambda$11287, and $\uplambda$1303 photons \citep{1947PASP...59..196B,1995ApJ...439..346K}. For V1741~Sgr, both of these O\,{\sc i} emission lines are strong, with a $\lambda$11287/$\lambda$8446 flux ratio of 1.5, moderately larger than the ratio of photon energies. The $\lambda$8446/Pa\,$\gamma$ ratio is 0.35, which yields a similar $\tau_{\mathrm{Pa}\gamma}=0.8$ as above \citep[][their Figure 15]{2011MNRAS.411.2383K}. For comparison, both LkH$\alpha$~225~S and Gaia~19ajj exhibit $\uplambda$11287 emission, but the $\uplambda$8446 line is in absorption in the former and not detected in the latter. \citet{2011MNRAS.411.2383K} note that $\uplambda$8446 absorption can result from radiative trapping of the $\uplambda$1303 photons, which may indicate optically thicker gas in those outbursters than in V1741~Sgr.  

\begin{figure*}
    \includegraphics[width=0.98\textwidth]{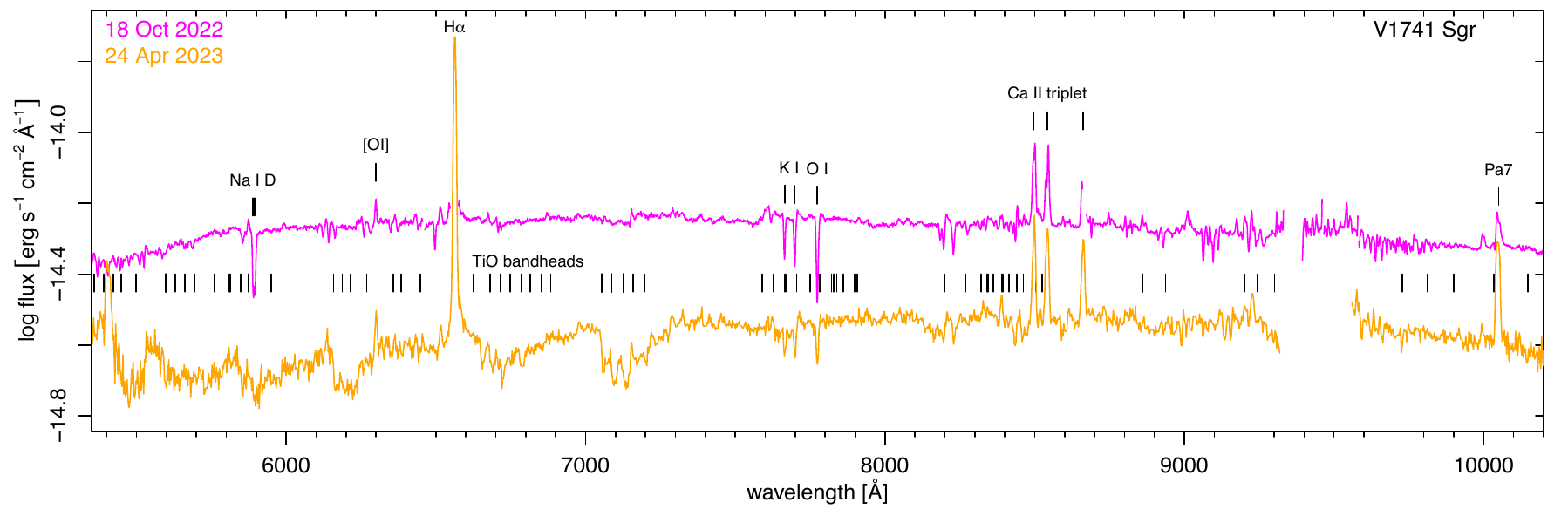}
    \caption{``Red'' side spectra during the outburst plateau (LRIS; 18 Oct 2022) and during the decline (Kast; 24 Apr 2023). Each spectrum is shown on an absolute, calibrated log-flux scale. Wavelengths of strong TiO bandheads are marked by the black tics as indicated.}
    \label{fig:spec-red}
\end{figure*}

\begin{figure}
    \includegraphics[width=0.48\textwidth]{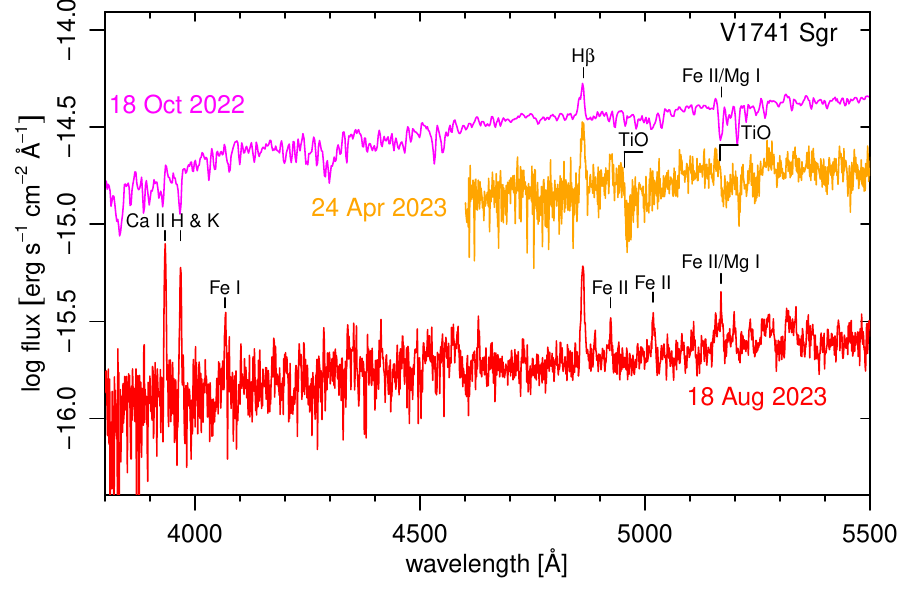}
    \caption{``Blue'' side spectra during the outburst plateau (LRIS; 18 Oct 2022), during the decline (Kast; 24 Apr 2023), and after return to quiescence (18 Aug 2023; Kast). Each spectrum is shown on an absolute, calibrated log-flux scale. Prominent lines and molecular bands are marked. A noisy region of the 24 Apr 2023 Shane/KAST spectrum, resulting from a shorter exposure time, is masked for visual clarity.}
    \label{fig:spec-blue}
\end{figure}

\begin{figure}
    \includegraphics[width=0.48\textwidth]{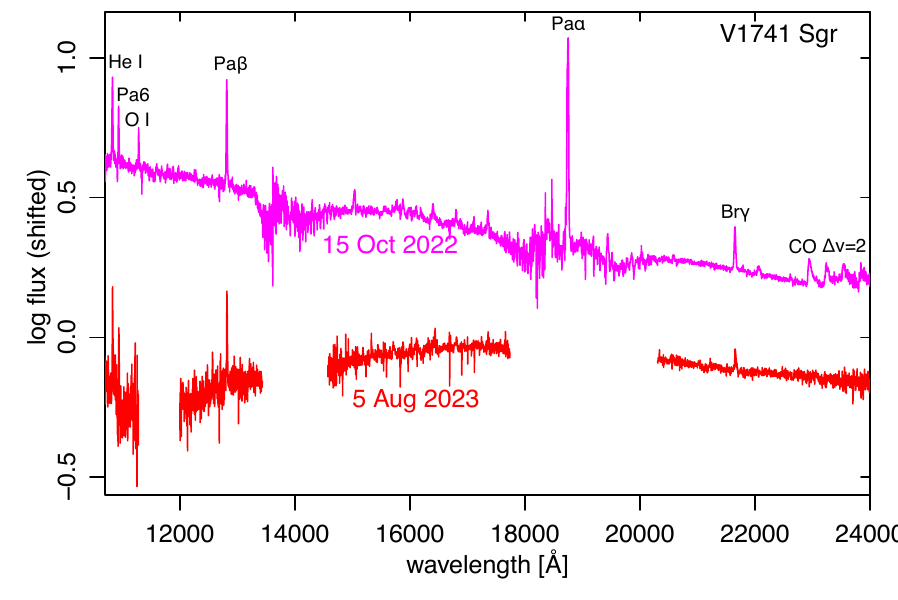}
    \caption{Near infrared spectra during the outburst plateau (SpeX; 15 Oct 2022) and after return to quiescence (TripleSpec; 5 Aug 2023). The TripleSpec spectrum does not have absolute flux calibration, so the relative vertical positioning of the spectra is arbitrary. }
    \label{fig:tspec}
\end{figure}

\section{Spectral Evolution}\label{sec:evolution}

Further monitoring of V1741~Sgr revealed spectroscopic changes to the source as it faded during the spring and summer of 2023. The first Shane/Kast spectra (24 April 2023) were taken after the source had declined about halfway from its peak. Nevertheless, even at this stage, the source was $\sim$5--10 times brighter than its pre-outburst state, with its flux still dominated by the outburst. However, by the next Shane/Kast observation (18 Aug 2023), the source had returned to its pre-outburst flux. Only the blue-side Kast spectrum was obtained on this date. A post-outburst TripleSpec observation (5 Aug 2023) was also made around the same time. 

The spectra in Figures~\ref{fig:spec-red}--\ref{fig:spec-blue} are presented with absolute flux calibration, revealing the decline in continuum flux over the full wavelength range with each subsequent spectrum. 

\subsection{Shane/Kast Spectra of the Decaying Outburst}

Between the LRIS observation and the first Shane/Kast observation, the absorption features dramatically changed, with TiO absorption becoming dominant as the source began to fade (Figures~\ref{fig:spec-red} and \ref{fig:spec-blue}). In the blue-side Kast spectrum, TiO bands appeared, starting at 4954~\AA\ and 5167~\AA\ (part of the TiO $\alpha$ system). On the red side, the TiO absorption features were even clearer (owing to the better signal-to-noise), with strong features starting at 5759~\AA\ ($\alpha$ system), 6149~\AA\ ($\gamma^\prime$ system), 6651~\AA, and 7045~\AA\ ($\gamma$ system). 

The pattern of TiO band strengths, plus the lack of detected VO absorption, suggests an early M spectral type according to the classification scheme of \citet{Kirkpatrick1991}. The spectral indices from \citet[][their Table~3]{Herczeg2014} provide quantitative spectral type estimates. For the April 2023 spectrum, the index $x_\mathrm{TiO7140} = -0.48$ indicates a spectral type of M1--M2 ($T\approx3560$--3720~K). The level of veiling in the $R$ band may be a source of systematic error on this estimate, with more veiling leading to earlier estimated spectral types \citep{Fang2020}. 

Several wind-associated absorption features weakened or disappeared. For example, the equivalent widths of the $\uplambda\uplambda$7665/7699 K\,{\sc i} lines and the $\uplambda$7774 O\,{\sc i} triplet decreased by a factor of $\sim$0.8 and the Na\,{\sc i}\,D doublet (blended with TiO) disappeared.

Conversely, the prominent emission lines do not weaken relative to the continuum but, in some cases, become stronger. Slight strengthening relative to the continuum is detected in H$\beta$, [O\,{\sc i}], and the Ca\,{\sc ii} triplet, and significant strengthening is detected in Pa7. In absolute flux, all these lines dimmed somewhat, apart from Pa7, but not as rapidly as the overall dimming of the source. 

\subsection{Post-Outburst Spectra}\label{sec:post}

The final Shane/Kast blue-side spectrum from August 2023 (Figure~\ref{fig:spec-blue}) again revealed changes in spectroscopic features as the star continued to dim. The TiO absorption, seen in the previous Kast blue-side spectrum, is gone. The absence of these features implies a spectral type of K9 or earlier. Furthermore, the MgH band, detected in late K stars, is also not detected, implying a spectral type of K5 or earlier \citep{GrayCorbally2009}. Together, these constraints indicate a spectroscopic temperature of $>$4140~K \citep{2013ApJS..208....9P}, 
implying that the emission was dominated by hotter gas at this epoch than the previous one. 

At this epoch, several emission lines appeared or strengthened.
 H$\beta$ continued to strengthen relative to the continuum. Furthermore, strong emission was seen from the Ca\,{\sc ii} H and K lines, which were absent during the outburst. This spectrum also shows the appearance of iron emission lines, including $\uplambda$4064 Fe\,{\sc i}, $\uplambda$4924 and $\uplambda$5018 Fe\,{\sc ii}, and the $\uplambda$5169 Fe\,{\sc ii}/Mg\,{\sc i} blend, with the latter having transitioned from absorption to emission.

The near-infrared also presented changes in spectroscopic features between 
the outburst and post-outburst epochs (Figure~\ref{fig:tspec}). Notably, emission was no longer detected from the CO bands, providing further evidence that the EX Lup-like outburst was over. H\,{\sc i} and He\,{\sc i} lines were still detected in this wavelength range, but the Pa$\beta$ and longer-wavelength hydrogen lines became weaker relative to the continuum, opposite the behaviour of hydrogen lines at shorter wavelengths.

\begin{figure*}
    \includegraphics[width=1\textwidth]{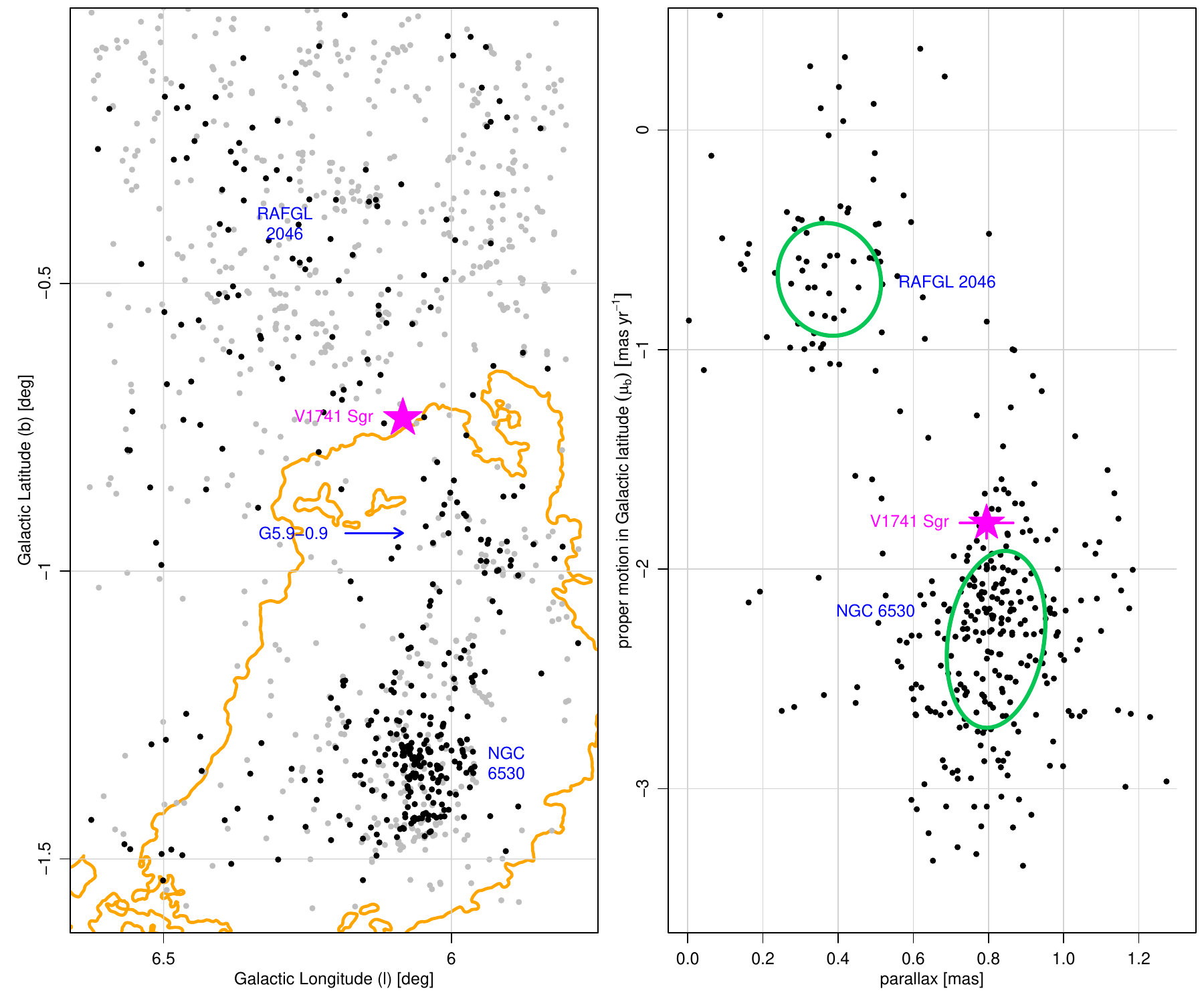}
    \caption{
        Left: Spatial distribution of YSOs in the neighbourhood of V1741~Sgr. YSOs with good Gaia astrometry are black points, those without are grey points, and V1741~Sgr is the magenta star. The three nearby YSO groups are labelled in blue. The optical nebulosity from the Lagoon Nebula is outlined with the orange contours derived from the Digitized Sky Survey red-band image \citep{Lasker1990}. 
        Right: Astrometry for V1741~Sgr (magenta star) and neighbouring YSOs (black points). The green ellipses mark the two Gaussian mixture-model components found in parallax--proper-motion space, one corresponding to the Lagoon Nebula and the other to RAFGL~2046. V1741~Sgr's astrometry is most consistent with the former.
    }
    \label{fig:mem}
\end{figure*}

\section{Stellar Properties} \label{sec:star}

The source of the outburst is likely a relatively isolated Classical T Tauri star, with a mass slightly less than our Sun and a relatively high quiescent accretion rate. The evidence for this characterisation, inferred from pre-outburst data, is presented in the following section.

\subsection{Lagoon Nebula Membership} \label{sec:mem}

V1741~Sgr is projected near the edge of the Lagoon Nebula, separated by $\sim$0.7$^\circ$ from the main NGC~6350 cluster \citep{Walker1957}. To verify whether this association is physical, we examined Gaia's Early Data Release 3  \citep[EDR3;][]{GaiaCollaboration2021}, which provides a 6-parameter astrometric solution for  V1741 Sgr \citep{Lindegren2021_astrom}. Its parallax is $\varpi=0.795\pm0.072$~mas, corrected for a zero-point offset of $\mathrm{zp}=-0.042$~mas \citep{Lindegren2021_zp}, which corresponds to a distance of $\sim$1260~pc. Its proper motions are $\mu_{\alpha^\star} = 1.360\pm0.076$~mas~yr$^{-1}$ and $\mu_\delta = -1.213\pm0.056$~mas~yr$^{-1}$ in equatorial coordinates, corresponding to  $\mu_{\ell^\star} = -0.379\pm0.061$~mas~yr$^{-1}$ and $\mu_b = -1.789\pm0.0716$~mas~yr$^{-1}$ in Galactic coordinates. The source's renormalised unit weight error of 1.033 is close to unity, it has a small non-statistically significant astrometric excess noise of 0.22~mas, and 16 visibility periods were used.

The projected locations of YSOs in the neighbourhood of V1741~Sgr were obtained from the SPICY catalogue (Figure~\ref{fig:mem}). In this catalog, candidate YSOs were identified solely from infrared photometry, employing spectral energy distribution (SED) fitting and statistical classification, meaning that these sources trace the distribution of star formation over a large swath of the Galactic midplane with minimal spatial bias \citep{SPICY}. In the vicinity of V1741~Sgr, three groups are detected:
\begin{description}
\item[\bf{G6.0-1.3}] This is the main star cluster in the Lagoon Nebula, including stars from NGC~6530 and the Hour Glass Nebula. Its mean parallax is $\varpi = 0.821\pm0.006$~mas ($d=1220_{-20}^{+30}$~pc), and its mean proper motions are $\mu_{\ell^\star} = -1.18$~mas~yr$^{-1}$ and $\mu_{b}=-2.19$~mas~yr$^{-1}$ \citep{Kuhn2021_sgr}.
\item[\bf{G5.9-0.9}] This group is half a degree north-west of NGC~6530, but lies within the boundary of the Lagoon Nebula. 
\citet{Kuhn2021_sgr} calculated a parallax of $0.774\pm0.020$~mas ($d=1290\pm30$~pc) and proper motions of $\mu_{\ell^\star,0} = -0.83$~mas~yr$^{-1}$ and $\mu_{b,0}=2.49$~mas~yr$^{-1}$ that are statistically consistent with the other Lagoon Nebula members. 
\item[\bf{G6.1-0.3}] This group corresponds RAFGL~2046. Its parallax, $\varpi_0=0.357\pm0.019$~mas \citep{Kuhn2021_sgr}, places it behind Lagoon at a distance of $2800_{-140}^{+150}$~pc. Its mean proper motions are $\mu_{\ell^\star,0} = -0.98$~mas~yr$^{-1}$ and $\mu_{b,0}=-0.68$~mas~yr$^{-1}$, with a $b$ component significantly discrepant from the Lagoon Nebula.
\end{description}

V1741~Sgr is projected between the latter two groups. However, the star's Gaia EDR3 parallax is most consistent with the Lagoon Nebula. We modelled the parallax--proper-motion distribution of the YSOs in this region with a mixture of Gaussians using the {\tt mclust} software \citep{mclust02,mclust5}. The best-fitting model included one component for the Lagoon Nebula stars (G6.0-1.3 and G5.9-0.9), another for RAFGL~2046 (G6.1-0.3), and a small non-clustered population of unassociated stars (Figure~\ref{fig:mem}, right). This model supports V1741~Sgr's Lagoon membership (97\% probability), with a lower propability of non-association (3\%), and a negligible probability of RAFGL~2046 membership ($<$0.001\%).

Within the Lagoon Nebula, the cluster NGC~6530 is expanding with an average outward velocity of $\sim$1~km~s$^{-1}$ \citep{2019ApJ...870...32K,2019MNRAS.486.2477W}. However, neither V1741~Sgr nor the nearby subcluster G5.9-0.9 are moving away from NGC~6530, as would be expected if they were to follow the same expansion pattern. This implies that these peripheral association members were not ejected from the central cluster but likely formed on the outskirts \citep[see similar cases in][]{2022ApJ...937...46K,2023AJ....165....3K}.

\subsection{Quiescent YSO Properties}\label{sec:pre}

\begin{figure}
    \includegraphics[width=0.48\textwidth]{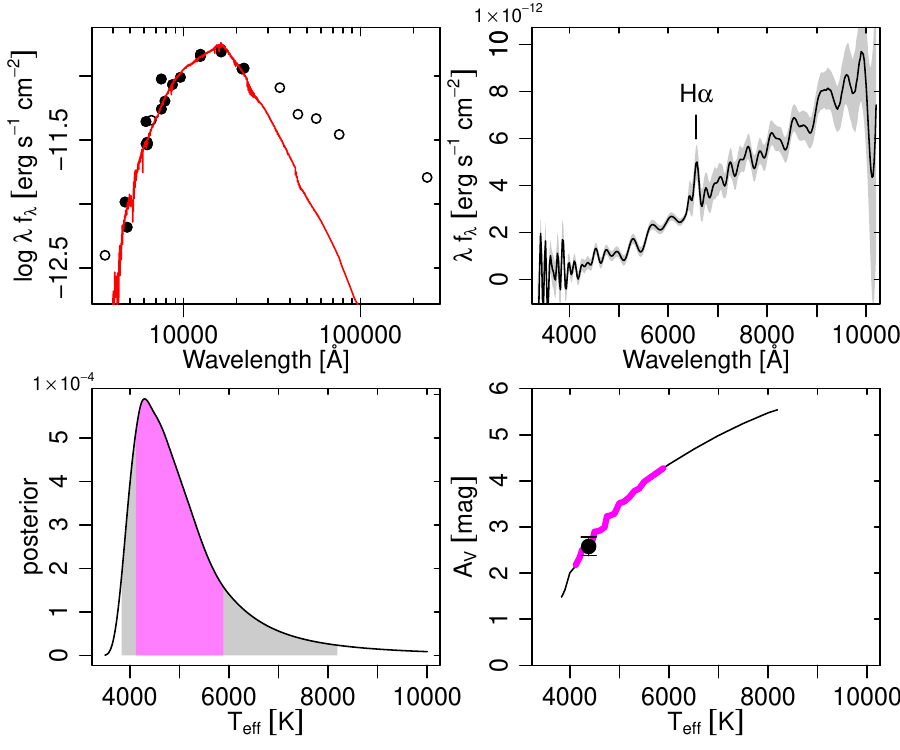}
    \caption{Pre-outburst properties of V1741~Sgr. Top left: The stellar SED. Photometric measurements used in the fit are black points, and those with excess emission are open circles. The maximum-likelihood stellar atmosphere model ($T_\mathrm{eff} \sim 4380$~K; $A_V \sim 2.6$~mag) is indicated by the red line. 
    Top right: The Gaia BP/RP spectrum, with the uncertainty envelope in grey. 
    Bottom left: The posterior distribution for $T_\mathrm{eff}$. The 68\% credible interval is shaded magenta, and the 95\% credible interval is shaded grey. 
    Bottom Right: The line illustrates degeneracy between $A_V$ on $T_\mathrm{eff}$ from SED fitting. The black point indicates the maximum-likelihood parameters, the error bar indicates uncertainty on $A_V$ for a fixed $T_\mathrm{eff}$, and the credible intervals are indicated by the magenta (68\%) and black (95\%) portions of the line.
    }
    \label{fig:pre}
\end{figure}

V1741~Sgr's pre-outburst SED suggests a reddened stellar photosphere with excess in the near-ultraviolet and mid-infrared bands (Figure~\ref{fig:pre}, top left). The time-averaged Gaia BP/RP spectrum for 2014--2017 \citep{2022arXiv220606143D,2022arXiv220606205M} shows a prominent H$\alpha$ emission line (Figure~\ref{fig:pre}, top right). These attributes are expected in the standard magnetospheric model of accreting T Tauri stars, where accretion luminosity provides the ultraviolet excess, disc emission provides the infrared excess, and H$\alpha$ emission is largely attributed to irradiated gas in accretion streams or winds \citep{2016ARA&A..54..135H}. 

To constrain the star's effective temperature ($T_\mathrm{eff}$) and $V$-band extinction ($A_V$), we fit the SED with BOSZ stellar atmosphere models \citep{Bohlin2017} reddened by the \citet{Cardelli89} extinction law in the optical/near-infrared and the \citet{Wang2015} law in the mid-infrared. The models were convolved with filter profiles obtained from the Spanish Virtual Observatory\footnote{\url{http://svo2.cab.inta-csic.es/theory/fps/}} \citep{2012ivoa.rept.1015R,2020sea..confE.182R} to predict flux in individual photometric bands. Our statistical model includes normally distributed photometric measurement errors ($\sigma_\mathrm{phot}$) and an additional normally distributed error term ($\sigma_\varepsilon$) to account for astrophysical scatter (e.g., stellar variability). Only photometric measurements without blue or infrared excess (i.e., those between 4600 and 22,000~\AA) were included in the fit. The log-likelihood function was maximised using the BFGS algorithm \citep{BFGS_B, BFGS_F, BFGS_G, BFGS_S} implemented by the $R$ function {\tt optim} \citep{RCore2022}. To calculate Bayesian posterior distributions, we assumed uniform priors for all parameters, with $3500\leq T_\mathrm{eff
}\leq10,000$~K, $0\leq A_V \leq 10$, $-10 \leq \log \sigma_\varepsilon \leq 10$, and $-100 \leq \log \mathrm{scale} \leq 0$. 

\begin{figure}
    \includegraphics[width=0.48\textwidth]{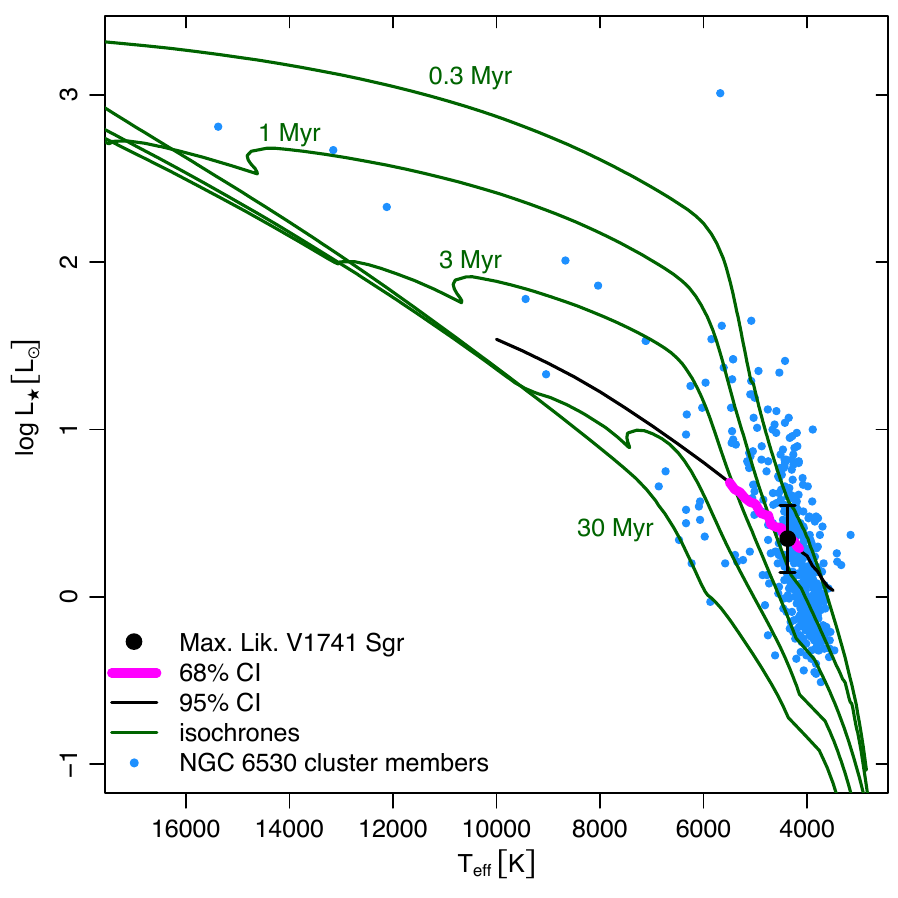}
    \caption{Hertzsprung-Russell diagram for NGC~6530. The plot shows the maximum-likelihood parameters for V1741~Sgr (black point), uncertainty on its $L_\star$ for a fixed $T_\mathrm{eff}$ (error bar), and its 68\% (magenta curve) and 95\% (black curve) credible intervals given the $A_V$--$T_\mathrm{eff}$ degeneracy. Cluster members with Gaia-ESO spectroscopy are indicated by the blue points. 
    Pre--main-sequence isochrones for 0.3--30~Myr \citep{Bressan2012} are indicated by the green curves. 
    }
    \label{fig:hrd}
\end{figure}

The maximum likelihood $T_\mathrm{eff}$ is 4380~K, which is consistent with the constraints from the post-outburst spectrum (Section~\ref{sec:post}). However, the posterior distribution has a heavy tail extending to higher temperatures, meaning that higher temperatures cannot be absolutely excluded (Figure~\ref{fig:pre}, bottom left). The 1$\sigma$ (68\%) credible interval is 4100--5850~K (i.e., spectral types K6--G2), whilst the 95\% credible interval is 3850--8200~K. Assuming a temperature of 4380~K, the corresponding extinction would be $A_V = 2.6\pm0.2$~mag. However, there is considerable degeneracy between $A_V$ and $T_\mathrm{eff}$ (Figure~\ref{fig:pre}, bottom right). Nevertheless, the models yield a lower limit of $A_V > 1.5$~mag, with possible extinctions extending to 5.5~mag for the highest temperature models. The $T_\mathrm{eff} = 4380$~K solution corresponds to a star with a luminosity (excluding contributions from accretion or disc emission) of $\log (L_\star /L_\odot)= 0.35\pm0.20$ and a radius of $R_\star =  2.6\pm0.3$~$R_\odot$. (The full range of possible luminosities are indicated on Figure~\ref{fig:hrd}.)

Our best-estimate pre-outburst $A_V$ of 2.6~mag is higher than the best-estimate $A_V\approx1.4$~mag during the outburst (Section~\ref{sec:extinction}), suggesting that extinction may have decreased during the outburst as has been observed in several other EX Lup-type events \citep[e.g.,][]{Lorenzetti2012}. Nevertheless, the broad posterior distributions for $A_V$ mean that this change is not highly statistically significant. For the remainder of the analysis, we will assume our best-estimate $A_V$ values. However, if $A_V$ did not change, the increase in brightness would need to be driven entirely by temperature and surface area increases.

V1741~Sgr lies in the pre-main--sequence region of the Hertzsprung-Russell diagram (Figure~\ref{fig:hrd}), along with other young members of the Lagoon Nebula whose temperatures and luminosities have been obtained from the Gaia-ESO survey \citep{Jackson2022}. Assuming that stars in the Lagoon Nebula have similar ages, the other members can serve as a guide to the most probable ($T_\mathrm{eff}$, $\log L$) combination for V1741~Sgr. Our best-fit parameters lie amid the densest group of low-mass Gaia-ESO cluster members, suggesting that they are reasonable estimates. These parameters correspond to a stellar mass of 0.8~$M_\odot$ and an age of 0.7~Myr in the PARSEC evolutionary models \citep{Bressan2012}. Nevertheless, the uncertainty in $T_\mathrm{eff}$ and the systematic differences between theoretical evolutionary models \citep[e.g.,][]{David2019,Braun2021} can cause estimates to vary by a factor of several.

The H$\alpha$-line flux from the calibrated Gaia BP/RP spectrum is $(2.5\pm0.6)\times10^{-14}$~erg~s$^{-1}$~cm$^{-2}$, corresponding to an equivalent width of $-52\pm14$~\AA. Assuming $A_V=2.6$~mag, the extinction-corrected line-luminosity would be $(3.4\pm0.5)\times10^{31}$~erg~s$^{-1}$. Applying the relation between H$\alpha$ and accretion luminosity from \citet{Alcala2014} gives 
\begin{equation}
\log (L_\mathrm{acc}/L_\odot) \approx 1.12 \times \log (L_{\mathrm{H}\alpha}/L_\odot) + 1.5 = -0.9.
\end{equation}
Then substituting a values of $M_\star = 0.8$~$M_\odot$ and $R_\star = 2.6$~$R_\odot$ into their Equation~1 yields an accretion rate of
\begin{equation}\label{eqn:mdot}
\dot{M} = \eta \frac{L_\mathrm{acc}R_\star}{GM_\star} = 2\times10^{-8}~M_\odot\,\mathrm{yr}^{-1},
\end{equation}
where $G$ is the gravitational constant and $\eta=1.25$ is an approximate correction factor for accreting material falling from a radius of $\sim$5~$R_\star$ \citep{Alcala2014}.
This $\dot{M}$ is at the high end for quiescent T Tauri stars \citep{2023ASPC..534..539M}. 

\section{Evolution of the SED}\label{sec:phys}

\subsection{Infrared Emission}\label{sec:ircolor}

\begin{figure}
    \includegraphics[width=0.5\textwidth]{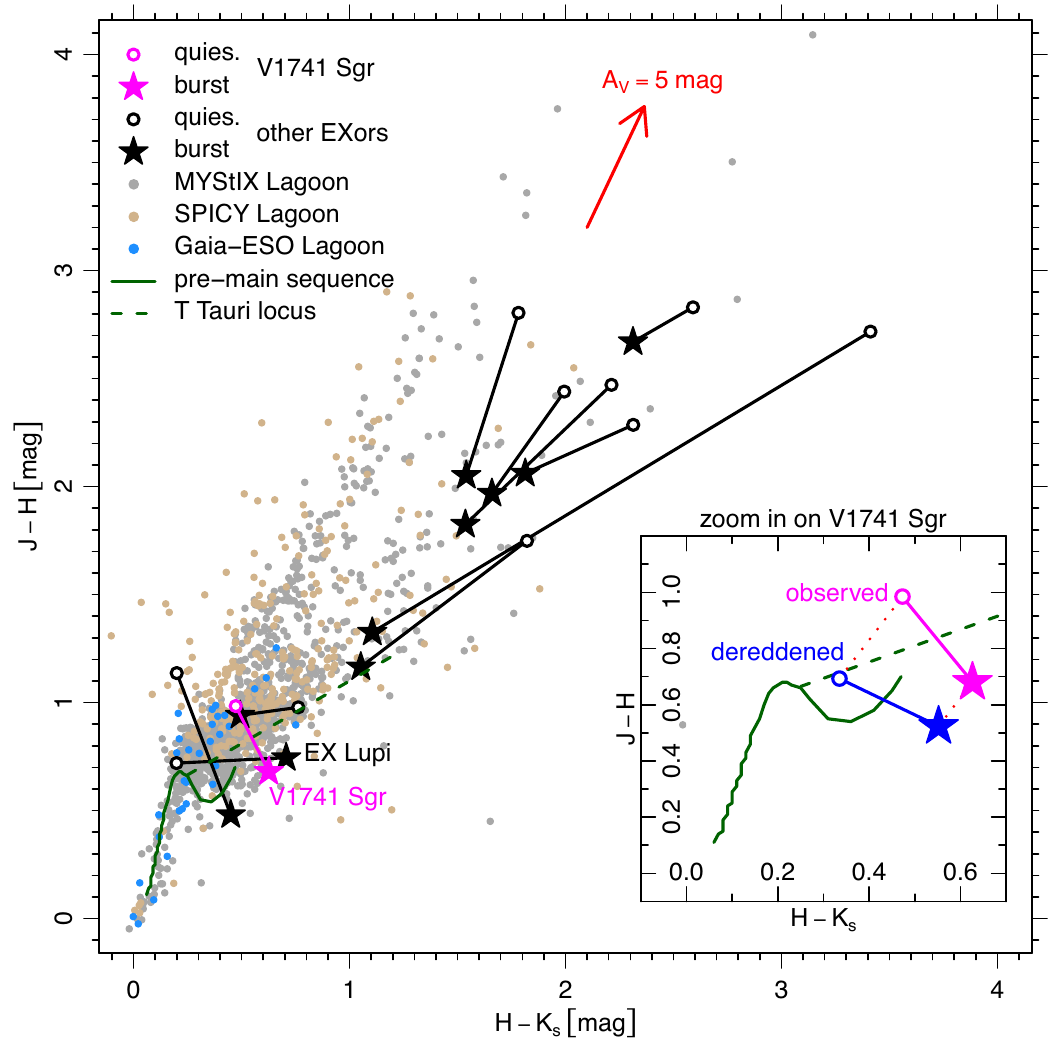}
    \caption{$J-H$ vs \ $H-K_s$ colours for YSOs, including V1741 Sgr (magenta), other EXors (black), and other members of the Lagoon Nebula region (grey: MYStIX; tan: SPICY; light blue: Gaia-ESO). For the outbursters, open circles indicate the quiescent $JHK_s$ colours and star symbols indicate outburst colours. The photometry for the other EXors was obtained from \citet[][their Table~3]{Lorenzetti2012}. (Note that the sources in their table differ from those in their Figure~1. The $JHK_s$ colours we show are calculated from the flux densities in their table, which we have calibrated to the VVV system.) The plot also indicates intrinsic colours of pre--main-sequence stars \citep{2013ApJS..208....9P}, the T Tauri locus \citep{1997AJ....114..288M}, and the reddening vector \citep{Wang2019}. The inset (bottom right) shows an expanded view of the observed (magenta) and dereddened (blue) colours of V1741 Sgr.
    }
    \label{fig:jhk}
\end{figure}

Infrared photometry can constrain whether brightening is driven by changes in extinction or luminosity (Figure~\ref{fig:jhk}). \citet{Lorenzetti2012} showed that many EX Lup-type outbursts became substantially bluer in both $J-H$ and $H-K$ colours, with colour changes too extreme to be solely temperature effects. Nevertheless, not all EX Lup-type stars in their sample demonstrated this behaviour -- notably EX Lup itself became redder in $H-K$ during its 2008 outburst but maintained a nearly constant $J-H$ colour. The NEOWISE light curves for V1741~Sgr showed the source becoming redder during the outburst by $\Delta(W1-W2)=0.3\pm0.1$~mag (Section~\ref{sec:outburst}). This is opposite what would be expected from decreasing extinction, implying that something else is occurring. 

V1741~Sgr's mean pre-outburst magnitudes of $J=13.8$~mag, $H=12.8$~mag, and $K_s=12.3$~mag were obtained from VVV. The SED fit in Section~\ref{sec:pre} indicates that the $J$ and $H$ bands are dominated by photospheric emission, whilst the $K_s$ band exhibits some infrared excess. Assuming a pre-outburst extinction of $A_V=2.6$~mag, the dereddened photometry is $J_\mathrm{dered} = 13.2$~mag, $H_\mathrm{dered} = 12.5$~mag, and $K_{s,\mathrm{dered}} = 12.1$~mag.

To characterise the near-infrared changes during the outburst, we generated synthetic photometry from the SpeX spectrum, using the VISTA $JHK_s$ filter profiles. This yielded values of $J=12.0$~mag, $H=11.3$~mag, and $K_s=10.7$~mag. Assuming an outburst extinction of $A_V = 1.4$~mag, the dereddened magnitudes are $J_\mathrm{dered}=11.7$~mag, $H_\mathrm{dered}=11.1$~mag, and $K_{s,\mathrm{dered}}=10.6$~mag. These correspond to a factor of 3.6--4 increase in luminosity in this near-infrared spectral region, meaning that the accretion luminosity dominates over the stellar photospheric emission at all wavelengths observed. 

On the $J-H$ vs.\ $H-K_s$ colour-colour diagram (Figure~\ref{fig:jhk}), the outburst shifted the star to the lower right, implying that the transition from becoming bluer during outburst to becoming redder during outburst happened around $\sim$1.6~$\upmu$m ($H$ band). The star started on the T Tauri locus from \citet{1997AJ....114..288M}, but this shift brought the star below the line. This behaviour differs from most of the sources examined by \citet{Lorenzetti2012}, which tend to get bluer in both colours. However, a few sources from the \citet{Lorenzetti2012} sample with bluer initial colours also defy this trend. A possible explanation is that the bluer sources start less embedded, limiting the impact of changes in extinction.

The dereddened outburst colours of $(J-H)_\mathrm{dered}=0.52$~mag and $(H-K_s)_\mathrm{dered}=0.55$~mag are similar, indicating that the outbursting source's SED has a nearly constant spectral index of 
\begin{equation}
\alpha = \frac{\mathrm{d}\log \lambda f_\lambda}{\mathrm{d}\log \lambda} \approx 1.7\,(J - K_{s})_\mathrm{dered} - 2.6 = -0.7
\end{equation}
in the near-infrared. When comparing to stellar atmosphere models from the PHOENIX library \citep{Husser2013}, the outburst's intrinsic $J-H$ colour corresponds to $T_\mathrm{eff} = 4400$~K for their $\log g=0$ models,  $T_\mathrm{eff} = 4600$~K for their $\log g=1$ models, or $T_\mathrm{eff} = 4700$~K for their $\log g=2$ models. These temperature constraints on the outbursting source are consistent with the temperature ranges estimated from the absorption features detected in the optical and near-infrared. 

\begin{figure}
    \includegraphics[width=0.5\textwidth]{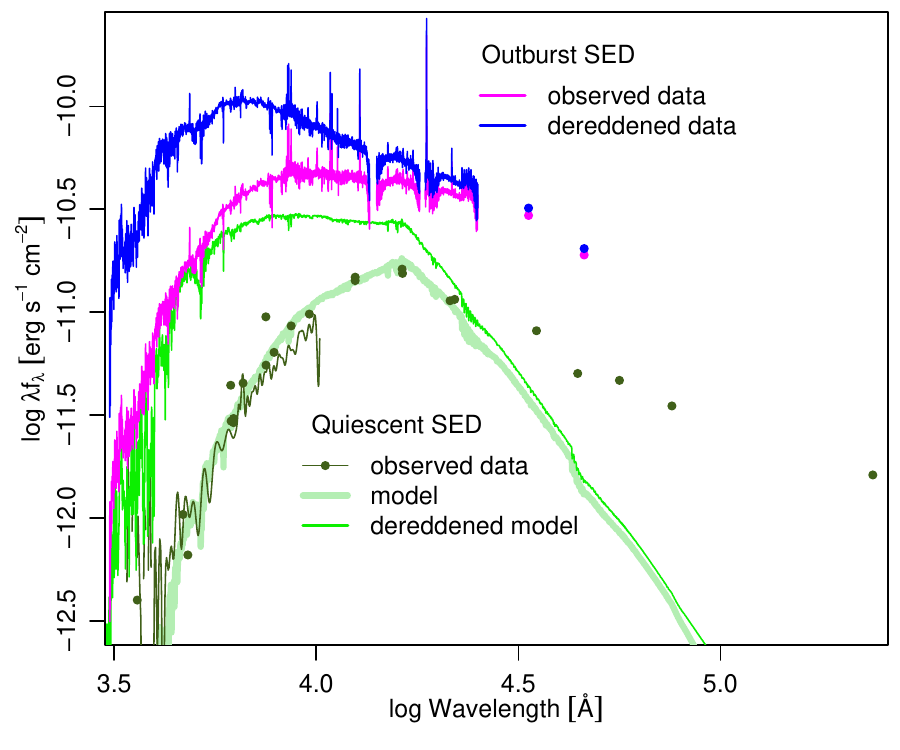}
    \caption{SEDs for V1741~Sgr before and during the outburst, with dereddening corrections applied, assuming $A_V=2.6$~mag in quiescence and $A_V=1.4$~mag during the outburst. The pre-outburst SED is constructed from multi-survey photometry and the Gaia BP/RP spectrum, while the outburst SED is constructed from the LRIS+SpeX spectroscopy and NEOWISE.
    }
    \label{fig:sed_compare}
\end{figure}

\subsection{Accretion Luminosity}\label{sec:acclum}

Examination of the SEDs of V1741~Sgr revealed that the source brightened over the full wavelength range in both observed and dereddened flux. The smallest increase was around $\sim$1.6~$\upmu$m ($H$ band), where a factor of $\sim$3.6 increase in the flux density was observed after applying the dereddening corrections (Figure~\ref{fig:sed_compare}). Assuming our best-fit stellar and outburst parameters, a factor of $\sim$2 increase in emitting stellar surface area would be needed to explain this $H$-band brightening. Additionally, a larger surface area during the outburst decline is indicated by TiO absorption features (April 2023), whose depth in flux units is significantly greater than the stellar continuum from the post-outburst (August 2023) spectrum.

The outburst's total accretion luminosity may be estimated by integrating the dereddened outburst SED and subtracting the stellar luminosity. We integrated this SED (blue curve in Figure~\ref{fig:sed_compare}) over the observed range, from $\sim$3000~\AA\ to 46,000~\AA\, using linear interpolation where there are gaps in the data. Although these integration limits do not include portions of the ultraviolet and infrared, the intrinsic spectrum is decreasing in both extremes, suggesting that the omitted spectral regions only provide small contributions to the bolometric luminosity. The accretion luminosity calculated in this way is $L_\mathrm{acc}=5$~$L_\mathrm{\odot}$.

Using Equation~\ref{eqn:mdot} and assuming the stellar parameters from Section~\ref{sec:pre}, we calculate a peak outburst accretion rate of $\dot{M}\approx6\times10^{-7}$~$M_\mathrm{\odot}$~year$^{-1}$. If we integrate the accretion over the full duration of the outburst, assuming that the accretion rate is proportional to the $r$-band flux increase, the total mass accreted is $6\times10^{-7}$~$M_\mathrm{\odot}$, or 0.2~$M_\oplus$.

The calculations above assumed $A_V$ estimates that indicate a slight decrease during the outburst (Section~\ref{sec:pre}). Here, we briefly examine the alternative scenario that $A_V$ did not change, remaining constant at $A_V=1.4$~mag. This would imply slightly larger changes in both temperature and dereddened $H$ band flux (see Sections~\ref{sec:post}, \ref{sec:pre}, and \ref{sec:ircolor}). These approximately compensate for each other, so the estimated change in emitting surface area would still be a factor of $\sim$2. 


\section{Discussion}\label{sec:discussion}

In broad terms, EX Lup-type outburst spectra resemble scaled-up T Tauri spectra \citep{Fischer2023}. However, the detection of low-gravity absorption lines from Ba\,{\sc ii} and Sr\,{\sc ii} (Section~\ref{sec:spec}) may suggest a difference in the structure of V1741~Sgr's emitting regions compared to typical T Tauri stars.

Explaining V1741~Sgr's continuum brightening also requires a change in emission structure during outburst.
In the standard T Tauri model, accretion excess (e.g., veiling) comes from hotspots on the stellar surface. However, hotspots are insufficient to explain the brightening of V1741~Sgr, which requires an emitting area $\sim$2 times larger than the stellar surface (Section~\ref{sec:acclum}). 
The impossibility of achieving the brightening via hotspots is underlined by the change in the outburst spectral type from K to early M during the beginning of the decline. When the early-M-type spectrum was observed, the star was still a factor of several brighter than its quiescent level, meaning that the spectrum was still dominated by the outburst and not the star (Figure~\ref{fig:spec-red}). However, the $\sim$3600~K implied by the spectrum is much cooler than expected from accretion hotspots. Furthermore, the early-M spectral type is cooler than the final post-outburst stellar spectrum, so, to outshine the star, this gas must have had a larger surface area (Figure~\ref{fig:spec-blue}). We can also rule out reprocessed emission from the disc or disc wall as the main contributor to the outburst continuum because dust produces a featureless spectrum, not the evolving absorption spectrum witnessed. We propose the alternative that the spectrum is dominated by circumstellar gas, which also explains the low gravity features.

The puzzling appearance then disappearance of the TiO bands during the transition of the star from outburst to quiescence (Figures~\ref{fig:spec-red} and \ref{fig:spec-blue}) can be explained by the above scenario. At the outburst peak, the circumstellar gas would have a photospheric temperature of $\sim$4750~K. This gas would start to cool during the outburst decay, leading to the cooler spectrum. Eventually, the circumstellar gas would fade sufficiently for the stellar spectrum, with a higher temperature, to reemerge.

In FU Ori outbursts, the luminosity is expected to originate from viscously heated discs. However, models based on a disc with a \citet{1988ApJ...325..231K} temperature profile cannot reproduce V1741~Sgr's SED. Using an implementation of this model by \citet{2022ApJ...927..144R}, it is impossible to simultaneously fit the blue and near-infrared portions of the source owing to the model's overestimation of the near-infrared emission. This is not surprising, given that the EX Lup phenomenon is generally attributed to changes near the star's magnetosphere rather than an increase in mass accretion through the whole disc \citep{SiciliaAguilar2012}. Given that the gas contributing to the increased continuum emission has a surface area a factor of several times the star's surface area, this gas could be distributed within either the stellar magnetosphere or in a small portion of the innermost disc. 

To investigate whether it is plausible that the increased continuum emission resulted from optically thick infalling gas in the magnetosphere, we performed a rough optical depth calculation. In our toy model, gas falls onto the star with an accretion rate $\dot{M}=6\times10^{-7}$~$M_\odot$~yr$^{-1}$, from a radius $r=5R_\star$, with a free-fall time of $t \approx ({\pi^2 r^3}/{8 G M_\star})^{1/2} \approx 1$~day. We further assume that this gas has a surface area (in projection) twice that of the star ($A \approx 2\pi R_\star^2$). The mean surface density of this material will be $\Sigma \approx \dot{M}tA^{-1} \sim 16$~g~cm$^{-2}$. The opacity of 4750~K gas is $\kappa \approx 0.001$--$0.01$~cm$^2$~g$^{-1}$ depending on volume density \citep[{\AE}SOPUS\footnote{\url{http://stev.oapd.inaf.it/cgi-bin/aesopus}} 2.0;][]{Marigo2009,Marigo2022}. This suggests that, if the infalling gas were distributed over the required surface area, it would be moderately optically thin and incapable of producing the increased continuum emission. Furthermore, the gas would need a temperature structure that generates the observed absorption spectrum. Nevertheless, the significant uncertainties in the toy model mean that such a scenario cannot be definitively ruled out. However, emission from the innermost portion of the gas disc seems a more likely explanation for the increased optical/near-infrared continuum. 

To explain the changing mid-infrared spectral index, it may be possible to invoke reprocessed emission alone. In the pre-outburst SED (Section~\ref{sec:pre}), the mid-infrared is dominated by reprocessed emission, whilst the near-infrared is dominated by stellar emission. The total reprocessed luminosity of a dust disc will scale proportionally to the total luminosity of the central illuminating source. As V1741~Sgr brightened, the emission of the central source shifted to be dominated by the blue part of the spectrum. If the mid-infrared reprocessed emission increases proportionally to the luminosity of the central source, it would increase faster than the near-infrared, which now contributes a smaller fraction to the total source luminosity. This effect could explain why  V1741~Sgr became redder beyond 1.6~$\upmu$m. The H$^-$ continuum opacity minimum around 1.6~$\upmu$m \citep{gray2005} also affects the relative temperature sensitivity of $H$-band flux, but this is likely to have been a minor effect owing to the coincidental similarity of the pre- and peak-outburst temperature estimates. Detailed modelling (beyond the scope of this paper) would be necessary to determine if a configuration of circumstellar material can reproduce the specific mid-infrared SED changes undergone by V1741~Sgr during this event.  

V1741~Sgr's relatively simple light curve, with onset, plateau, and decay phases lasting several months each, can be explained by the accretion scenario for EX Lup developed by \citet{SiciliaAguilar2012}. In this scenario, accretion through stable channels in the magnetosphere would yield the rapid onset and recovery of the system and the relatively constant brightness during the high state. V1741 Sgr's light curve is simpler than most other objects in its class, many of which have repeated bursts and outbursts of varying amplitudes or exhibit significant substructure in their outburst light curves \citep[e.g.,][]{2010ApJ...719L..50A,Ninan2015,2022ApJ...941..165P}. However, the small burst seen by PGIR in the month before the main outburst suggests that the outburst began in a gradual and choppy way. This would be consistent with mass release due to accretion instability \citep{2010MNRAS.406.1208D,2012MNRAS.420..416D,2015arXiv150906382A}, but argues against triggering by a sudden event (e.g., a stellar flyby).

Comparison of V1741 Sgr to LkH$\alpha$ 225 S and Gaia~19ajj highlights the diversity of spectra of outbursting YSOs. Differences include V1741 Sgr's stronger H\,{\sc i} and He\,{\sc i} lines, stronger O\,{\sc i}, and its weakness or lack of many of the emission lines species like Fe\,{\sc i}, Mg\,{\sc i}, Ti\,{\sc i}, or K\,{\sc i}. Although CO emission is a defining feature of EX Lup-type outbursts, there can be significant differences between how the bandheads appear, with V1741 Sgr having relatively strong emission. Absorption lines with higher excitation potentials (C\,{\sc i}, S\,{\sc i}, P\,{\sc i}) are seen in the spectrum of V1741~Sgr, but not LkH$\alpha$ 225 S and Gaia~19ajj. Finally, LkH$\alpha$ 225 S shows some wavelength-temperature dependence, but V1741~Sgr does not.

\section{Conclusion}\label{sec:conclusion}

Our analysis identifies this year-long outburst from V1741~Sgr as an EX Lup-type event of moderate amplitude ($\sim$3~mag in the optical and 1-2~mag in the infrared). The star is a classical T Tauri star (Section~\ref{sec:pre}) on the outskirts of the Lagoon Nebula (Section~\ref{sec:mem}), and the outbursting spectrum is characterised by strong atomic (H\,{\sc i}, He\,{\sc i}, Ca\,{\sc ii}, and O\,{\sc i}) and molecular (CO) emission (Section~\ref{sec:spec}).  The spectrum reveals brightening across the entire observed range, from the near-ultraviolet to the mid-infrared, causing a partial flattening of the SED (Section~\ref{sec:acclum}). Other features common in EX Lup-type objects, such as deeper absorption from wind-sensitive lines and low-gravity absorption features, are also present in this source (Sections~\ref{sec:optical_spectrum} and \ref{sec:spex}). 

The spectral evolution of the outburst from its peak to its decline and, finally, its post-outburst state provides clues about the origin of emission from EX Lup-type sources. The most notable change in V1741~Sgr's spectrum was the appearance and subsequent disappearance of TiO absorption while it continued to fade (Section~\ref{sec:evolution}). The timing of the appearance of the TiO absorption implies that this feature is connected to the outburst rather than the stellar photosphere, supporting our conclusion that circumstellar emission dominates the outburst continuum spectrum (Section~\ref{sec:discussion}). 

Our observations help build the sample of spectroscopically characterised EX Lup-type events. However, a detailed comparison of spectroscopic features reveals numerous differences between objects (Section~\ref{sec:spec}), suggesting that accretion discs and stellar magnetospheres have a variety of ways to interact. Although outbursts are most common from embedded protostars \citep{Lorenzetti2012,Audard2014,2017MNRAS.465.3011C,2017MNRAS.465.3039C}, those from more evolved YSOs, including V1741~Sgr and the prototypical EX Lup, help us understand the later stages of disc evolution, especially during phases when outbursts may affect forming planets. 

The basic properties of the star and the outburst are summarised below:
\begin{itemize}
\item V1741 Sgr was formed on the outskirts of the Lagoon Nebula, near a subcluster designated G5.9-0.9 \citep{Kuhn2021_sgr}, rather than the main NGC~6530 cluster. The star's Gaia parallax indicates a distance of $\sim$1260~pc. Environmental extinction includes $A_V=0.6$~mag from the foreground and $\Delta A_V=0.6$~mag from a cloud at the star's distance (Sections~\ref{sec:extinction} and \ref{sec:mem}).
\item The pre-outburst SED suggests that the star has an early-to-mid K spectral type and $A_V\sim2.6$~mag of extinction. This extinction is higher than that of the cloud, indicating a significant intrinsic component. Both near-ultraviolet and mid-infrared excesses are detected. We derive an age $\lesssim$3~Myr and a quiescent accretion rate of $\sim2\times10^{-8}$~$M_\odot$~yr$^{-1}$ (Section~\ref{sec:pre}).
\item The outburst light curve consists of a $\sim$60~day rise, a $\sim$280~day plateau, then a $\sim$110~day decay. The rise followed a sigmoid-like function, with slight lags in the rises at longer wavelengths. The decay was more choppy, with multiple temporary dips (Section~\ref{sec:outburst}). 
\item In the final photometric epochs, after the end of the outburst, the source dimmed to a fainter state than its pre-outburst level.
\item The outburst caused the SED to become bluer at wavelengths $<$1.6~$\upmu$m and redder at wavelengths $>$1.6~$\upmu$m (Section~\ref{sec:ircolor}). The increased blueness is likely a combination of increased temperature and decreased extinction, whilst the increased redness is likely an increase in reprocessed emission. 
\item There may have been a moderate decrease in extinction during the outburst ($\Delta A_V \sim -1.2$~mag). Nevertheless, this decrease is insufficient to explain the outburst's colour or brightness in either the optical (Section~\ref{sec:color}) or infrared (Section~\ref{sec:ircolor}). Stochastic variability before the outburst and during the plateau is consistent with variable extinction.
\item In the LRIS spectrum (Oct 2022), absorption features indicate a $\sim$K2 spectral type ($\sim$4750~K),
with no evidence for a wavelength-temperature dependence (Sections~\ref{sec:optical_spectrum} and \ref{sec:spex}). The strengths of low-gravity absorption features (Ba\,{\sc ii} and  Sr\,{\sc ii}) suggest that these lines were not formed on the stellar surface. 
\item Sequential spectroscopic observations revealed an evolving spectral type (Section~\ref{sec:evolution}). In the first Kast spectra (April 2023), absorption features, especially TiO, indicated an M1--M2 spectral type ($\sim$3600~K). In the final Kast spectrum (August 2023), the disappearance of TiO absorption and the absence of MgH implied a spectral type of K5 or earlier ($T \gtrsim 4140$~K). 
\item Analysis of the H emission lines with the \citet{2011MNRAS.411.2383K} model indicates the additional presence of optically thin $\sim$5000~K gas (Section~\ref{sec:kf}). 
\item The relative strengths of H\,{\sc i}, He\,{\sc i}, Ca\,{\sc ii}, O\,{\sc i}, Fe\,{\sc i}, Mg\,{\sc i}, Ti\,{\sc i}, Na\,{\sc i}, K\,{\sc i}, and CO features differ notably between several compared EX Lup-like outbursts. Strong Fe emission, which is often seen from accreting T Tauri stars, only emerged once the outburst faded.
\item The accretion luminosity in the source's high state is $\sim$5~$L_\odot$, corresponding to an outburst accretion rate of $\sim6\times10^{-7}$~$M_\odot$~yr$^{-1}$. At this rate, the event would have deposited 0.2~$M_\oplus$ onto the star (Section~\ref{sec:acclum}).
\end{itemize}

Combining large YSO catalogues with sky-monitoring surveys will likely improve the count statistics of these events. For example, as of December 2023, there have been $>$300 Gaia alerts for sources in the SPICY catalogue, including other possible EX Lup-like events \citep[e.g.,][]{2023RNAAS...7...57K}. The growing lists of YSO outbursts will likely fill in the spectroscopic parameter space and improve our understanding of typical properties and how these events vary. 

\section*{Acknowledgements}

We thank Calum Morris and Zhen Guo for valuable comments, Leigh Smith for access to proprietary VVV data products, and Facundo P\'erez Paolino for participation in the Palomar observation. We also thank the referee for thorough and helpful reviews of this paper. We acknowledge the significant cultural role and reverence that the summit of Maunakea has within the indigenous Hawaiian community and that we are most fortunate to have the opportunity to conduct observations from this mountain. IRTF is operated by the University of Hawaii under contract 80HQTR19D0030 with NASA. The W.\ M.\ Keck Observatory is operated as a scientific partnership among the California Institute of Technology, the University of California, and NASA. The Observatory was made possible by the generous financial support of the W.\ M.\ Keck Foundation. The Kast Double Spectrograph was made possible by a generous gift from William and Marina Kast. We acknowledge ESA Gaia, the Data Processing and Analysis Consortium, and the Photometric Science Alerts Team. ZTF is led by the California Institute of Technology, US, supported by the National Science Foundation under Grant No. AST-2034437, and includes IPAC, US; Los Alamos National Laboratory, US; University of Maryland, US; the University of Wisconsin at Milwaukee, US; University of Washington, US; Oskar-Klein Center of the University of Stockholm, Sweden; DESY and Humboldt University of Berlin, Germany; Weizmann Institute of Science, Israel; and the University System of Taiwan, Taiwan. 

For the purpose of open access, the authors have applied a Creative Commons Attribution (CC BY) license to any Author Accepted Manuscript version arising from this submission.

\section*{Data Availability}

The spectra from Table~\ref{tab:obslog} and the TAGRA and PGIR light curves are provided online as supplementary material. Other datasets are publicly available from the sources in Section~\ref{sec:data}.

\bibliographystyle{mnras}
\bibliography{ms.bbl} 

\bsp	
\label{lastpage}
\end{document}